\documentclass{article}                                                                           

\usepackage[margin=1.3in]{geometry}                                                     

\usepackage{setspace}                                                                             
\usepackage{indentfirst}                                                                          
\renewcommand\thesection{\Roman{section}}                                         
\renewcommand\thesubsection{\thesection.\Alph{subsection}}                   
\renewcommand\thesubsubsection{\thesubsection.\arabic{subsubsection}} 

\usepackage[explicit]{titlesec}                                                                  
\titleformat{\section}{\bfseries}{\thesection.}{1em}{\MakeUppercase{#1}}  
\titleformat{\subsection}{\bfseries}{\thesubsection.}{1em}{#1}                   
\titleformat{\subsubsection}{\itshape}{\thesubsubsection.}{1em}{#1}          

\usepackage{lastpage}                                                                             
\usepackage[figure,table]{totalcount}                                                        

\usepackage{amsmath}                                                                            

\usepackage{tablefootnote}
\usepackage{scrextend}

\usepackage[]{footmisc}                                                                          

\usepackage{graphicx}                                                                             
\usepackage{multirow}                                                                             

\usepackage{caption}                                                                              
\usepackage[labelformat=simple]{subcaption}                                           
\usepackage[colorlinks=true, allcolors=blue]{hyperref}                               

\usepackage{amssymb}
\def\R{\mathbb{R}}
\def\E{\mathbb{E}}
\newtheorem{rem}{\bf \it Remark}[section]

\usepackage{xcolor}
\usepackage{soul}

\begin{document}


\title{Advanced methodology for uncertainty propagation in computer experiments with large number of inputs} 

\author{
\vspace{20mm}
\\Bertrand Iooss,$^{\text{a,b},\ast}$ and Amandine Marrel,$^{c}$ \\[4pt] 
\textit{$^a$EDF R\&D, D\'epartement PRISME} \\ [-10pt]
\textit{6 Quai Watier, 78401 Chatou, France} \\ [-5pt]
\textit{$^b$Institut de Math\'ematiques de Toulouse} \\ [-10pt]
\textit{31062 Toulouse, France} \\ [-5pt]
\textit{$^c$CEA, DEN, DER, SESI, LEMS}\\[-10pt]       
\textit{13108 Saint-Paul-lez-Durance, France} \\[-2pt]
{$^\ast$Email: \href{mailto:bertrand.iooss@edf.fr}{bertrand.iooss@edf.fr}}}       


\clearpage\maketitle
\thispagestyle{empty}

\pagebreak
~\vfill

\begin{abstract}
In the framework of the estimation of safety margins in nuclear accident analysis, a quantitative assessment of the uncertainties tainting the results of computer simulations is essential. Accurate uncertainty propagation (estimation of high probabilities or quantiles) and quantitative sensitivity analysis may call for several thousand of code simulations. Complex computer codes, as the ones used in thermal-hydraulic accident scenario simulations, are often too cpu-time expensive to be directly used to perform these studies. A solution consists in replacing the computer model by a cpu inexpensive mathematical function, called a metamodel, built from a reduced number of code simulations. 
However, in case of high dimensional experiments (with typically several tens of inputs), the metamodel building process remains difficult. 
To face this limitation, we propose a methodology which combines several advanced statistical tools: initial space-filling design, screening to identify the non-influential inputs, Gaussian process (Gp) metamodel building with the group of influential inputs as explanatory variables. The residual effect of the group of non-influential inputs is captured by another Gp metamodel. Then, the resulting joint Gp metamodel is used to accurately estimate Sobol' sensitivity indices and high quantiles (here $95\%$-quantile).
The efficiency of the methodology to deal with a large number of inputs and reduce the calculation budget is illustrated on a thermal-hydraulic calculation case simulating with the CATHARE2 code a Loss Of Coolant Accident scenario in a Pressurized Water Reactor. A predictive Gp metamodel is built with only a few hundred of code simulations and allows the calculation of the Sobol' sensitivity indices. This Gp also provides a more accurate estimation of the $95\%$-quantile and associated confidence interval than the empirical approach, at equal calculation budget. Moreover, on this test case, the joint Gp approach outperforms the simple Gp.

\vspace{1em}\noindent\textbf{Keywords} --- Gaussian process, Metamodel, Quantile, Screening, Sobol' indices
\end{abstract}

\vfill

\pagebreak
\section{Introduction}

Best-estimate computer codes are increasingly used to estimate safety margins in nuclear accident management analysis instead of conservative procedures \cite{poumod09,bucpet10}.
In this context, it is essential to evaluate the accuracy of the numerical model results, whose uncertainties come mainly from the lack of knowledge of the underlying physics and the model input parameters.
The so-called Best Estimate Plus Uncertainty (BEPU) methods were then developed and introduced in safety analyses, especially for thermal-hydraulic issues, and even more precisely for the large-break loss-of-coolant accident \cite{promav07,wil13,sansan18}.
Its main principles rely on a probabilistic modeling of the model input uncertainties, on Monte Carlo sampling for running the thermal-hydraulic computer code on sets of inputs, and on the application of statistical tools (based for example on order statistics as the Wilks' formula) to infer high quantiles of the output variables of interest \cite{nutwal04,wal07,decbaz08,petdau08,marnut11}.

Not restricted to the nuclear engineering domain, the BEPU approach is more largely known as the uncertainty analysis framework \cite{derdev08}.
Quantitative assessment of the uncertainties tainting the results of computer simulations is indeed a major topic of interest in both industrial and scientific communities.
One of the key issues in such studies is to get information about the output when the numerical simulations are expensive to run. 
For example, one often faces up with cpu time consuming numerical models and, in such cases, uncertainty propagation, sensitivity analysis, optimization processing and system robustness analysis become difficult tasks using such models.
In order to circumvent this problem, a widely accepted method consists in replacing cpu-time expensive computer models by cpu inexpensive mathematical functions (called metamodels) based, e.g., on polynomials, neural networks, or Gaussian processes \cite{fanli06}.
This metamodel is built from a set of computer code simulations and must be as representative as possible of the code in the variation domain of its uncertain parameters, with good prediction capabilities. 
The use of metamodels has been extensively applied in engineering issues as it provides a multi-objective tool \cite{forsob08}: once estimated, the metamodel can be used to perform global sensitivity analysis, as well as uncertainty propagation, optimization, or calibration studies.
In BEPU-kind analyses, several works \cite{cangar08,ziodim09,lorzan11} have introduced the use of metamodels and shown how this technique can help estimate quantiles or probability of failure in thermal-hydraulic calculations.

However, the metamodeling technique is known to be relevant when simulated phenomena are related to a small number of random input variables (see \cite{ghahig17} for example).
In case of high dimensional numerical experiments (with typically several tens of inputs), and depending on the complexity of the underlying numerical model, the metamodel building process remains difficult, or even impracticable.
For example, the Gaussian process (Gp) model \cite{sanwil03} which has shown strong capabilities to solve practical problems, has some caveats when dealing with high dimensional problems. 
The main difficulty relies on the estimation of Gp hyperparameters.
Manipulating pre-defined or well-adapted Gp kernels (as in  \cite{muerou12,durgin13}) is a current research way, while several authors propose to couple the estimation procedure with variable selection techniques \cite{welbuc92,marioo08,woolew17}.

In this paper, following the latter technique, we propose a rigorous and robust method for building a Gp metamodel with a high-dimensional vector of inputs.
Moreover, concerning the practical use of this metamodel, our final goal is twofold: to be able to interpret the relationships between the model inputs and outputs with a quantitative sensitivity analysis, and to have a reliable high-level quantile estimation method which does not require additional runs of the computer code.

In what follows, the system under study is generically denoted
\begin{equation}
Y=g\left(X_1,\ldots,X_d\right)
\end{equation}
where $g(\cdot)$ is the numerical model (also called the computer code), whose output $Y$ and input parameters $X_1,\ldots,X_d$ belong to some measurable spaces $\mathcal{Y}$ and $\mathcal{X}_1, \ldots, \mathcal{X}_d$ respectively. 
$\mathbf{X}=\left(X_1,\ldots,X_d\right)$ is the input vector and we suppose that $\mathcal{X}=\prod_{k=1}^d\mathcal{X}_k \subset  \R^d $ and $\mathcal{Y}\subset\R$. 
For a given value of the vector of inputs $\mathbf{x} = \left(x_1,\ldots,x_d\right) \in \mathbb{R}^d$, a simulation run of the code yields an observed value $y = g(\mathbf{x})$. 

To meet our aforementioned objectives, we propose a sequential methodology which combines several relevant statistical techniques. Our approach consists in four steps:
\begin{enumerate}

\item \textbf{Step 1: Design of experiments.} Knowing the variation domain of the input variables, a design of $n$ numerical experiments is firstly performed and yields $n$ model output values. The obtained sample of inputs/outputs constitutes the learning sample. The goal is here to explore, the most efficiently and with a reasonable simulation budget, the domain of uncertain parameters $\mathcal{X}$ and get as much information as possible about the behavior of the simulator output $Y$. For this, we use a space-filling design (SFD) of experiments, providing a full coverage of the high-dimensional input space \cite{fanli06,woolew17}.
  
\item \textbf{Step 2: Preliminary screening.} From the learning sample, a screening technique is performed in order to identify the Primary Influential Inputs (PII) on the model output variability and rank them by decreasing influence. To achieve it, we use dependence measures with associated statistical tests. These measures have several advantages: they can be directly estimated on a SFD, the sensitivity indices that they provide are interpretable quantitatively and their efficiency for screening purpose has been recently illustrated by \cite{dav15,delmar16b,ioomar17}. 

\item \textbf{Step 3: Building and validation of joint metamodel.} From the learning sample, a metamodel is built to fit the simulator output $Y$. For this, we propose to use a joint Gp metamodel \cite{marioo12}, by considering only the PII as the explanatory inputs while the other inputs (screened as non significantly influential) are considered as a global stochastic (\textit{i.e. unknown}) input. Moreover, we use a sequential process to build the joint Gp where the ranked PII are successively included  as explanatory inputs in the metamodel (ranking from Step 2). At each iteration, a first Gp model \cite{marioo08}, only depending on the current set of explanatory inputs, is built to approximate the mean component. The residual effect of the other inputs is captured using a second Gp model, also function of the explanatory inputs,
which approximates the variance component.
The accuracy and prediction capabilities of the joint metamodel are controlled on a test sample or by cross-validation. 

\item \textbf{Step 4: Use of the metamodel for sensitivity analysis and uncertainty propagation}. A quantitative sensitivity analysis (Step 4A) and an uncertainty propagation (Step 4B) are performed using the joint metamodel instead of the computer model, leading to a large gain of computation time \cite{zabdej98,marioo12}. In this work, we are particularly interested in estimating variance-based sensitivity indices (namely Sobol' indices) and the $95\%$-quantile of model output.
\end{enumerate}

\begin{figure}[!ht]
\centering
      \includegraphics[,height=16.5cm]{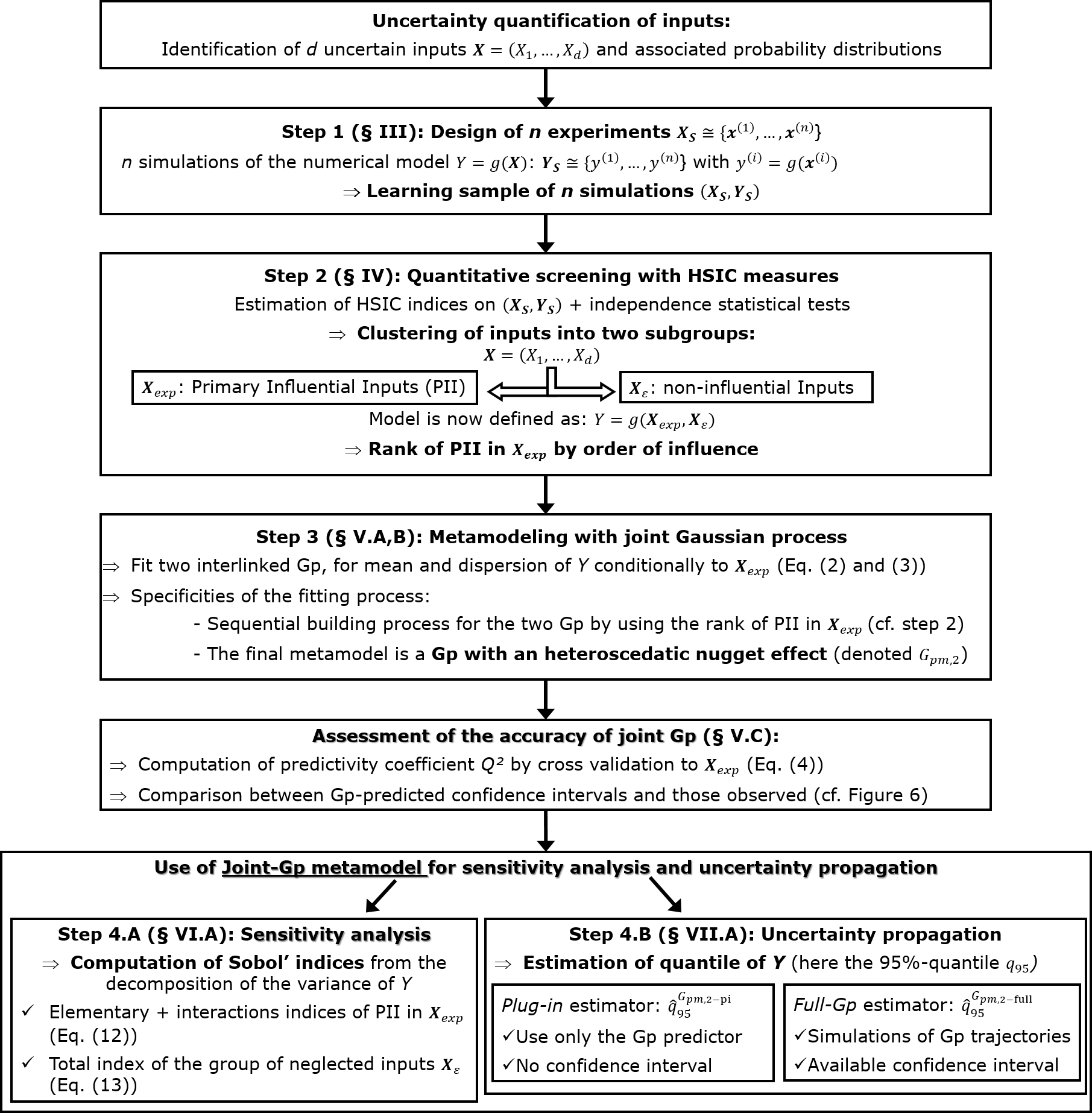}
\caption{\label{fig:Workflow} General workflow of the statitical methodology.}
\end{figure}

For ease of reading and understanding of the methodology, Figure \ref{fig:Workflow} gives a general workflow of the articulation of the main steps. For each step, the dedicated sections, the main notations that will be properly introduced later and the key equations are also referenced, in order to provide a guideline for the reader. 
The paper is organized as follows. In the following section, the nuclear test case which constitutes the guideline application of the paper is described. It consists in a thermal-hydraulic calculation case which simulates an accidental scenario in a nuclear Pressurized Water Reactor. Then, each step of the above  statistical methodology is detailed in a dedicated section and illustrated as the same time on the use-case. A last section concludes the work.

\section{Thermal-hydraulic test case}

\begin{figure}[!ht]
\centering
\includegraphics[,height=7cm]{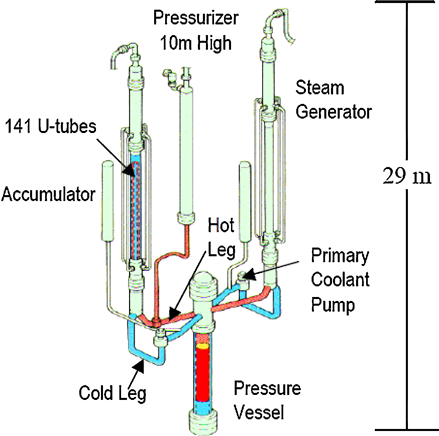}
\vspace{12pt}
\caption{\label{fig:LSTF}Large Scale Test Facility showing the main components and the hot and cold legs.}
\end{figure}

Our use-case consists in thermal-hydraulic computer experiments, typically used in support of regulatory work and nuclear power plant design and operation. 
Indeed, some safety analysis considers the so-called ``Loss Of Coolant Accident'' which takes into account a double-ended guillotine break with a specific size piping rupture.
The test case under study is a simplified one, as regards both physical phenomena and dimensions of the system, with respect to a realistic modeling of the reactor. 
The numerical model is based on code CATHARE2 (V2.5\_3mod3.1) which simulates the time evolution of physical quantities during a thermal-hydraulic transient. 
It models a test carried out on the Japanese mock-up ``Large Scale Test Facility'' (LSTF) in the framework of the OECD/ROSA-2 project, and which is representative of an Intermediate Break Loss Of Coolant Accident (IBLOCA) \cite{mazvac16}. 
This mock-up represents a reduced scale Westinghouse PWR ($1/1$ ratio in height and $1/48$ in volume), with two loops instead of the four loops on the actual reactor and an electric powered heating core ($10$ MWe), see Figure \ref{fig:LSTF}. 
It operates at the same pressure and temperature values as the reference PWR. The simulated accidental transient is an IBLOCA with a break on the cold leg and no safety injection on the broken leg.
The test under study reproduces a PWR $17\%$ (of cold leg cross-sectional area) cold leg IBLOCA transient with total failure of the auxiliary feedwater, single failure of diesel generators and three systems only available in the intact loop (high pressure injection, accumulator and low pressure injection).

\begin{figure}[!ht]
\centering
\includegraphics[,height=7cm]{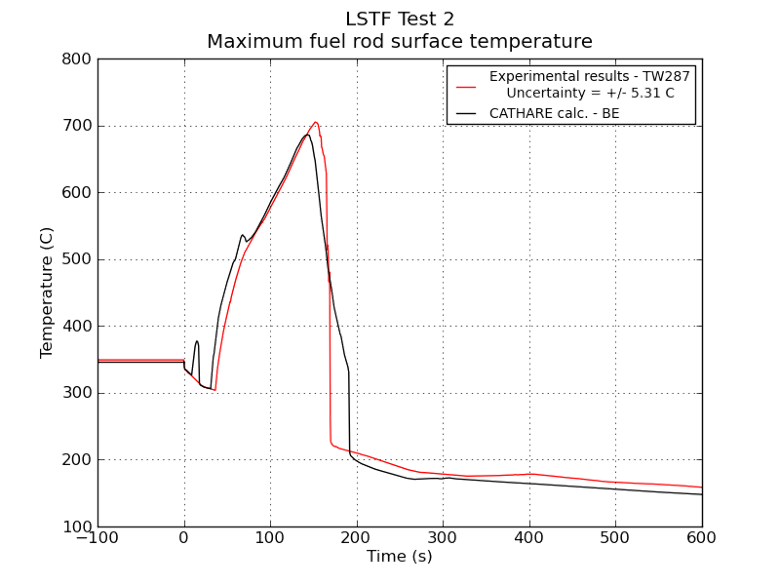}
\vspace{12pt}
\caption{Experimental values and physical simulation output of the CATHARE2 model: maximal rod cladding temperature during the transient.}\label{fig:temp}
\end{figure}

CATHARE2 is used to simulate this integral effect test (see \cite{geivac17} for the full details of the modeling).
During an IBLOCA, the reactor coolant system minimum mass inventory and the Peak Cladding Temperature (PCT) are obtained shortly after the beginning of the accumulators' injection.
Figure \ref{fig:temp} shows the CATHARE2 prediction and the experimental values of the maximal cladding temperature (also called maximal heater rod temperature) obtained during the test.
The conclusion of \cite{geivac17}, which also presents other results, is that the CATHARE2 modeling of the LSTF allows to reproduce the global trends of the different physical phenomena during the transient of the experimental test.
In our study, the main output variable of interest will be a single scalar which is the PCT during the accident transient (as an example, see the peak in Figure \ref{fig:temp}). 
This quantity is derived from the physical outputs provided by CATHARE2 code.

The input parameters of the CATHARE2 code correspond to various system parameters as boundary conditions, some critical flow rates, interfacial friction coefficients, condensation coefficients, heat transfer coefficients, etc. 
In our study, only uncertainties related to physical parameters are considered and no uncertainty on scenario variables (initial state of the reactor before the transient) is taken into account.
All uncertain physical models identified in a IBLOCA transient of a nuclear power plant are supposed to apply to the LSTF, except phenomena related to fuel behavior because of the fuel absence in the LTSF.
A physical model uncertainty consists in an additive or multiplicative coefficient associated to a physical model.
Finally, $d=27$ scalar input variables of CATHARE2 code are considered uncertain and statistically independent of each other.
They are then defined by their marginal probability density function (uniform, log-uniform, normal or log-normal).
Table \ref{tab:inputs} gives more details about these uncertain inputs and their probability density functions (pdf). 
The nature of these uncertainties appears to be epistemic  since  they  come  from a  lack of knowledge on the true value of these parameters. 

\vspace{0.3cm}

\begin{table}[!ht]
\caption{List of the $27$ uncertain input parameters and associated physical models in CATHARE2 code.}\label{tab:inputs}
\begin{tabular}{c|c|c|l}\hline
\bf Type of inputs & \bf Inputs & \bf pdf \tablefootnote{\samepage U, LU, N and LN respectively stands for Uniform, Log-Uniform, Normal and Log-Normal probability distributions.} & \bf Physical models  \\
\hline
Heat transfer & $X_1$ & N & Departure from nucleate boiling\\
in the core & $X_2$ & U & Minimum film stable temperature\\
& $X_3$ & LN & HTC\tablefootnote{\samepage Heat Transfer Coefficient.} for steam convection\\
& $X_4$ & LN & Wall-fluid HTC \\
& $X_5$  & N & HTC for film boiling \\\hline
Heat transfer in the steam & $X_6$ & LU & HTC forced wall-steam convection\\
generators (SG) U-tube & $X_7$ & N & Liquid-interface HTC for film condensation \\\hline
Wall-steam friction in core & $X_8$ & LU & \\\hline
Interfacial friction & $X_9$ & LN & SG outlet plena and crossover legs together\\
& $X_{10}$ & LN & Hot legs (horizontal part)\\
& $X_{11}$ & LN & Bend of the hot legs\\
& $X_{12}$ & LN & SG inlet plena\\
& $X_{13}$ & LN & Downcomer\\
& $X_{14}$ & LN & Core\\
& $X_{15}$ & LN & Upper plenum\\
& $X_{16}$ & LN & Lower plenum\\
& $X_{17}$ & LN & Upper head\\\hline
Condensation & $X_{18}$ & LN & Downcomer\\
& $X_{19}$ & U & Cold leg (intact)\\
& $X_{20}$ & U & Cold leg (broken)\\
& $X_{27}$ & U & Jet\\\hline
Break flow & $X_{21}$ & LN & Flashing (undersaturated)\\
& $X_{22}$ & N & Wall-liquid friction (undersaturated)\\
& $X_{23}$ & N & Flashing delay (undersaturated)\\
& $X_{24}$ & LN & Flashing (saturated)\\
& $X_{25}$ & N & Wall-liquid friction (saturated)\\
& $X_{26}$ & LN & Global interfacial friction (saturated)\\\hline
\end{tabular}
\end{table}

Our objective with this use-case is to provide a good metamodel for sensitivity analysis, uncertainty propagation and, more generally, safety studies.
Indeed, the cpu-time cost of one simulation is too important to directly perform all the statistical analysis which are required in a safety study and for which many simulations are needed.
To overcome this limitation, an accurate metamodel, built from a reduced number of direct code calculations, will make it possible to develop a more complete and robust safety demonstration.

\section{Step 1: Design of experiments}\label{sec:SFD}

This initial step of sampling is to define a design of $n$ experiments for the inputs and performing the corresponding runs with the numerical model $g$. The obtained sample of inputs/outputs will constitute the learning sample on which the metamodel will then be fitted. The objective is therefore to investigate, most efficiently and with few simulations, the whole variation domain of the uncertain parameters in order to build a predictive metamodel which approximates as accurately as possible the output $Y$. 

For this, we propose to use a space-filling design (SFD) of a $n$ of experiments, this kind of design providing a full coverage of the high-dimensional input space \cite{fanli06}. Among SFD types, a Latin Hypercube Sample (LHS, \cite{mckbec79})) with optimal space-filling and good projection properties \cite{woolew17} would be well adapted.  In particular, \cite{fanli06,damcou13} have shown the importance of ensuring good low-order sub-projection properties. Maximum projection designs \cite{josgul15} or low-centered $L^2$ discrepancy LHS \cite{jinche05} are then particularly well-suited.

Mathematically, the experimental design corresponds to a $n$-size sample $\left\{ \mathbf{x}^{(1)},\ldots,\mathbf{x}^{(n)}\right\}$ which is performed on the model (or code) $g$. 
This yields $n$ model output values denoted $\left\{ y^{(1)},\ldots,y^{(n)} \right\}$ with $y^{(i)} = g(\mathbf{x}^{(i)})$. 
The obtained learning sample is denoted $\left( X_s, Y_s\right)$ with $X_s = \left[{\mathbf{x}^{(1)}}^T,\ldots,{\mathbf{x}^{(n)}}^T\right]^T$ and $Y_s = \left[y^{(1)},\ldots,y^{(n)}\right]^T$. 
Then, the goal is to build an approximating metamodel of $g$ from the $n$-sample $\left( X_s, Y_s\right)$. 

The number $n$ of simulations is a compromise between the CPU time required for each simulation and the number of input parameters. For uncertainty propagation and metamodel-building purpose, some rules of thumb propose to choose $n$ at least as large as $10$ times the dimension $d$ of the input vector \cite{loesac09,marioo08}.

\begin{figure}[!ht]
\centering
\includegraphics[,height=6cm]{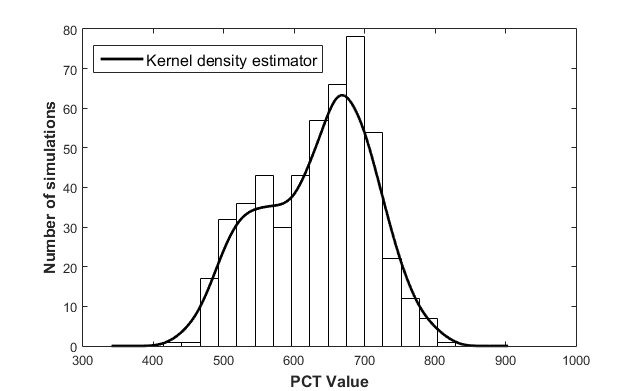}
\caption{\label{fig:hist}Histogram of the PCT from the learning sample of $n = 500$ simulations.}
\end{figure}

To build the metamodel for the IBLOCA test case, $n = 500$ CATHARE2 simulations are performed following a space-filling LHS built in dimension $d = 27$. The histogram of the obtained values for the output of interest, namely the PCT, is given by Figure \ref{fig:hist} (temperature is in $^{\circ}$C). 
A kernel density estimator \cite{par62} of the data is also added on the plot to provide an estimator of the probability density function. 
A bimodality seems to be present in the histogram.
It underlines the existence of bifurcation or threshold effects in the code, probably caused by a phenomenon of counter current flow limitation between the bend of hot legs and the steam generator inlet plena.

\begin{rem}
Note that the input values are sampled following their prior distributions defined on their variation ranges. Indeed, as we are not ensured to be able to build a sufficiently accurate metamodel, we prefer sample the inputs following the probabilistic distributions in order to have at least a probabilized sample of the uncertain output, on which statistical characteristics could be estimated. Moreover, as explained in the next section, dependence measures can be directly estimated on this sample, providing first usable results of sensitivity analysis. 
\end{rem}

Finally, the bimodality that has been observed on the PCT distribution strengthens the use of advanced sensitivity indices (i.e. more general than linear ones or variance-based ones) in our subsequent analysis.

\section{Step 2: Preliminary screening based on dependence measure}\label{sec:screening}

From the learning sample, a screening technique is performed in order to identify the primary influential inputs (PII) on the variability of model output. It has been recently shown that screening based on dependence measures \cite{dav15,delmar16b} or on derivative-based global sensitivity measures \cite{kucioo17,roubar17} are very efficient methods which can be directly applied on a SFD. Moreover, beyond the screening job, these sensitivity indices can be quantitatively interpreted and used to order the PII by decreasing influence, paving the way for a sequential building of metamodel.
In the considered IBLOCA test case, the adjoint model is not available and the derivatives of the model output are therefore not computed because of their costs. The screening step will then be based only on dependence measures, more precisely on Hilbert-Schmidt independence criterion (HSIC) indices, directly estimated from the learning sample.

The dependence measures for screening purpose have been proposed by \cite{dav15} and \cite{delmar16b}. 
These sensitivity indices are not the classical ones based on the decomposition of output variance (see \cite{ioolem15} for a global review). They consider higher order information about the output behavior in order to provide more detailed information. 
Among them, the Hilbert-Schmidt independence criterion (HSIC) introduced by \cite{grebou05} builds upon kernel-based approaches for detecting dependence, and more particularly on cross-covariance operators in reproducing kernel Hilbert spaces, see Appendix \ref{sec:HSIC} for mathematical details.

From the estimated $R^2_{\text{HSIC}}$ \cite{dav15}, independence tests can be performed for a screening purpose. 
The objective is to separate the inputs into two sub-groups, the significant ones and the non-significant ones.
For a given input $X_k$, statistical HSIC-based tests aims at testing the null hypothesis ``$\mathcal{H}_0^{(k)}$: $X_k$ and $Y$ are independent'', against its alternative ``$\mathcal{H}_1^{(k)}$: $X_k$ and $Y$ are dependent''. The significance level\footnote{\samepage The significance level of a statistical hypothesis test is the probability of rejecting the null hypothesis $\mathcal{H}_0$ when it is true.} of these tests is hereinafter noted $\alpha$. Several HSIC-based statistical tests are available: asymptotic versions based on an approximation with a Gamma law (for large sample size), spectral extensions and permutation-based versions for non-asymptotic case (case of small sample size). All these tests are described and compared in \cite{delmar16b}; a guidance to use them for a screening purpose is also proposed.

So, at the end of this HSIC-based screening step, the inputs are clustered into two subgroups, PII and non-influential inputs, and the PII are ordered by decreasing $R^2_{\text{HSIC}}$.
This order will be used for the sequential metamodel building in step 3.

On the IBLOCA test case, from the learning sample of $n = 500$ simulations, $R^2_{\text{HSIC}}$ dependence measures are estimated and bootstrap independence tests with $\alpha = 0.1$ are performed. The independence hypothesis $\mathcal{H}_0$ is rejected for eleven inputs, which are now designated as PII (Primary Influential Inputs). 
These inputs are given by Table \ref{tab:R2HSIC} with their estimated $R^2_{\text{HSIC}}$. 
Ordering them by decreasing  $R^2_{\text{HSIC}}$ reveals:
\begin{itemize}
\item the large influence of the interfacial friction coefficient in the horizontal part of the hot legs (X$_{10}$), 
\item followed by the minimum stable film temperature in the core X$_{2}$, the interfacial friction coefficient in the SG inlet plena X$_{12}$ and the wall to liquid friction (in under-saturated break flow conditions) in the break line X$_{22}$, 
\item followed by seven parameters with a lower influence: the interfacial friction coefficients in the upper plenum X$_{15}$, the downcomer X$_{13}$, the core X$_{14}$ and the SG outlet plena and crossover legs together X$_{9}$, the heat transfer coefficient in the core for film boiling X$_{5}$, the interfacial friction coefficient of the saturated break flow X$_{26}$ and the condensation coefficient in the jet during the injection X$_{27}$.
\end{itemize}
These results clearly underline the predominance influence of the uncertainties on various interfacial friction coefficients.

\vspace{0.3cm}

\begin{table}[!ht]
\caption{HSIC-based sensitivity indices $R^2_{\text{HSIC}}$ for the PII (Primary Influential Inputs), identified by independence test (Step 2).}\label{tab:R2HSIC}
\begin{tabular}{c|lllllllllll}\hline
PII &  X$_{10}$ & X$_{2}$ & X$_{12}$ & X$_{22}$ & X$_{15}$ & X$_{13}$ & X$_{9}$ & X$_{5}$ & X$_{14}$ & X$_{26}$ & X$_{27}$ \\
\hline
$R^2_{\text{HSIC}}$ & 0.39 & 0.04 & 0.02 & 0.02 & 0.01 & 0.01   & 0.01  & 0.01  & 0.01  & 0.01 &0.01
\end{tabular}
\end{table}

\begin{figure}[!ht]
      \includegraphics[width=5cm]{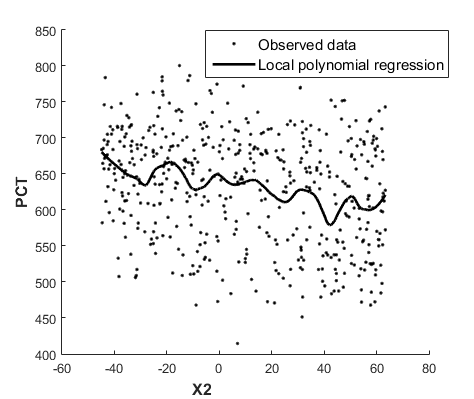}
      \includegraphics[width=5cm]{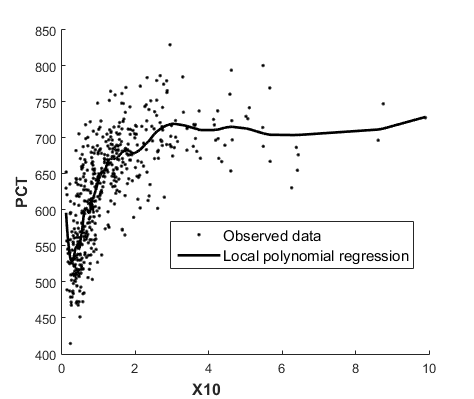}
     \includegraphics[width=5cm]{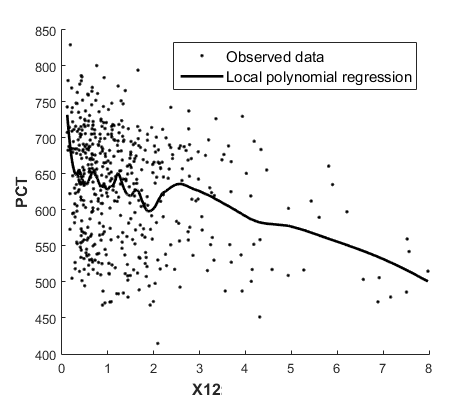}
\caption{\label{fig:scatter}Scatterplots with local polynomial regression of PCT according to several inputs, from the learning sample of $n = 500$ simulations.}
\end{figure}

From the learning sample, some scatterplots of the PCT with respect to some well-chosen inputs (the three most influential ones: X$_2$, X$_{10}$ and X$_{12}$) are displayed in Figure \ref{fig:scatter}. 
An additional local regression using weighted linear least squares and a first degree polynomial model (moving average filter) is added on each scatterplot to extract a possible tendency.
One can observe that larger values of the interfacial friction coefficient in the horizontal part of the hot legs (X$_{10}$) lead to larger values of the PCT.
This can be explained by the increase of vapor which brings the liquid in the horizontal part of hot legs, leading to a reduction of the liquid water return from the rising part of the U-tubes of the SG to the core (through the hot branches and the upper plenum). 
Since the amount of liquid water available to the core cooling is reduced, higher PCT are observed. 
In addition, we notice a threshold effect concerning this input: beyond a value of $2$, the water non-return effect seems to have been reached, and X$_{10}$ no longer appears to be influential.
We also note that the minimum stable film temperature in the core (X$_2$) shows a trend: the more it increases, the lower the PCT. This is explained by the fact that in the film boiling regime in the core (i.e. when the rods are isolated from the liquid by a film of vapor), X$_2$ represents (with a decrease in heat flux) the temperature from which the thermal transfer returns to the nucleate boiling regime. Thus, the larger X$_2$, the faster the re-wetting of the rods, the faster the cladding temperature excursion is stopped, and thus the lower the PCT.

The same kind of physical analysis can be made for other PII by looking at their individual scatterplot. 
Finally, it is important to note that the estimated HSIC and the results of significant tests are relatively stable when the learning sample size varies from $n = 300$ to $n = 500$. Only two or three selected variables with a very low HSIC ($R^2_{\text{HSIC}}$ around  0.01) can differ. This confirms the robustness, with respect to the sample size, of the estimated HSIC and the results of the associated significance tests. Their relevance for qualitative ranking and screening purpose is emphasized.

In the next steps, only the eleven PII are considered as explanatory variables in the joint metamodel and will be successively included in the building process. The other sixteen variables will be joined in a so-called \textit{uncontrollable} parameter.

\section{Step 3: Joint Gp metamodel with sequential building process}\label{seq:JM_building}

Among all the metamodel-based solutions (polynomials, splines, neural networks, etc.), we focus our attention on the Gaussian process (Gp) regression, which extends the kriging principles of geostatistics to computer experiments by considering the correlation between two responses of a computer code depending on the distance between input variables. The Gp-based metamodel presents some real advantages compared to other metamodels: exact interpolation property, simple analytical formulations of the predictor, availability of the mean squared error of the predictions and the proved capabilities for modeling of numerical simulators (see \cite{sacwel89}, \cite{sanwil03} or \cite{marioo08}). The reader can refer to \cite{raswil06} for a detailed review on Gp metamodel. 
 
However, for its application to complex industrial problems, developing a robust implementation methodology is required. Indeed, it often implies the estimation of several hyperparameters involved in the covariance function of the Gp (e.g. usual case of anisotropic stationary covariance function). Therefore, some difficulties can arise from the parameter estimation procedure (instability, high number of hyperparameters, see \cite{marioo08} for example). 
To tackle this issue, we propose a progressive estimation procedure based on  the result of the previous screening step and using a joint Gp approach \cite{marioo12}. 
The interest of the previous screening step becomes twofold. First, the input space, on which each component of the joint Gp is built, can be reduced to the PII space (only the PII are explanatory inputs of Gp). Secondly, the joint Gp is built with a sequential process where the ranked PII are successively included as explanatory inputs in the metamodel. It is expected that these two uses of screening
results could significantly make the joint Gp building easier and more efficient.

\subsection{Sequential Gp-building process based on successive inclusion of PII as explanatory variables}\label{methodo:sucess}

At the end of the screening step, the PII are ordered by decreasing influence (decreasing $R^2_{\text{HSIC}}$). They are successively included as explanatory inputs in the Gp metamodel while the other inputs (the remaining PII and the other non-PII inputs) are joined in a single macro-parameter which is considered as an uncontrollable parameter (i.e. a stochastic parameter, notion detailed in section \ref{jointGp}). 
Thus, at the $j^{th}$ iteration, a joint Gp metamodel is built with, as explanatory inputs, the $j$ first ordered PII. The definition and building procedure of a joint Gp is fully described in \cite{marioo12} and summarized in the next subsection.

However, building a Gp or a joint Gp involves to perform a numerical optimization in order to estimate all the parameters of the metamodel (covariance hyperparameters and variance parameter). As we usually consider in computer experiments anisotropic (stationary) covariance, the number of hyperparameters linearly increases with the number of inputs. In order to improve the robustness of the optimization process and deal with a large number of inputs, the estimated hyperparameters obtained at the $(j-1)^{th}$ iteration are used, as starting points for the optimization algorithm. This procedure is repeated until the inclusion of all the PII. Note that this sequential estimation process is directly adapted from the one proposed by \cite{marioo08}.  

\subsection{Joint Gp metamodeling}\label{jointGp}

We propose to use a joint Gp metamodeling to handle the group of non-PII inputs and capture their  residual effect.
In the framework of stochastic computer codes, \cite{zabdej98} proposed to model the mean and dispersion of the code output by two interlinked Generalized Linear Models (GLM), called ``joint GLM''.
This approach has been extended by \cite{marioo12} to several nonparametric models and the best results on several tests are obtained with two interlinked Gp models, called ``joint Gp''. 
In this case, the stochastic input is considered as an uncontrollable parameter denoted $\mathbf{X}_\varepsilon$ (i.e. governed by a seed variable). 

We extend this approach to a group of non-explanatory variables. More precisely, the input variables $\mathbf{X}=(X_1,\ldots,X_d)$ are divided in two subgroups: the explanatory ones denoted $\mathbf{X_{exp}}$ and the others denoted $\mathbf{X}_\varepsilon$.
The output is thus defined by $ Y = g(\mathbf{X_{exp}},\mathbf{X}_\varepsilon)$ and the metamodeling process will now focus on fitting the random variable $Y|\mathbf{X_{exp}}$\footnote{\samepage \label{Yrandom}$Y|\mathbf{X_{exp}}$ (i.e. $Y$ knowing $X_{exp}$) is a random variable as its value depends on the uncontrollable random variable $\mathbf{X}_\varepsilon$.}.
Under this hypothesis, the joint metamodeling approach yields building two metamodels, one for the mean $Y_m$ and another for the dispersion component $Y_d$:
\begin{equation}\label{eqYm}
  Y_m(\mathbf{X_{exp}}) = \mathbb{E}(Y|\mathbf{X_{exp}}) 
  \end{equation}
  \begin{equation}\label{eqYd}
  Y_d(\mathbf{X_{exp}}) = \mbox{Var}(Y|\mathbf{X_{exp}}) = \mathbb{E}\left[ (Y-Y_m(\mathbf{X_{exp}}))^2 |\mathbf{X_{exp}} \right]. 
\end{equation}
where $\mathbb{E}[Z]$ is the usual notation for the expected (i.e. mean) value of a random variable $Z$.

To fit these mean and dispersion components, we propose to use the methodology proposed by \cite{marioo12}. To summarize, it consists in the following steps. First, an initial Gp denoted $Gp_{m,1}$ is estimated for the mean component with homoscedastic nugget effect\footnote{\samepage Borrowed from geostatistics to refer to an unexpected nugget of gold found in a mining process, a constant nugget effect assumes an additive white noise effect, whose variance constitutes the nugget parameter. Most often, this variance is assumed to be constant, independent from the inputs (here $X_{exp}$), and the nugget effect is called homoscedastic. When this variance depends on the value of $x$ (i.e. is a function of $X$), the nugget effect is called heteroscedastic.}. A nugget effect is required to relax the interpolation property of the Gp metamodel. Indeed, this property, which would yield zero residuals for the whole learning sample, is no longer desirable as the output $Y|\mathbf{X_{exp}}$ is stochastic. Then, a second Gp, denoted $Gp_{v,1}$, is built for the dispersion component with, here also, an homoscedastic nugget effect. $Gp_{v,1}$ is fitted on the squared residuals from the predictor of $Gp_{m,1}$. Its predictor is considered as an estimator of the dispersion component. The predictor of $Gp_{v,1}$ provides an estimation of the dispersion at each point, which is considered as the value of the heteroscedastic nugget effect. The homoscedastic hypothesis is so removed and a new Gp, denoted $Gp_{m,2}$, is fitted on data, with the estimated heteroscedastic nugget. This nugget, as a function of $\mathbf{X_{exp}}$, accounts for a potential interaction between $\mathbf{X_{exp}}$ and the uncontrollable parameter $\mathbf{X}_\varepsilon$. Finally, the Gp on the dispersion component is updated from $Gp_{m,2}$ following the same methodology as for $Gp_{v,1}$. 

\begin{rem}
Note that some parametric choices are made for all the Gp metamodels: a constant trend and a Mat\'{e}rn stationary anisotropic covariance are chosen. All the hyperparameters (covariance parameters) and the nugget effect (when homoscedastic hypothesis is done) are estimated by maximum likelihood optimization process.
\end{rem}

\subsection{Assessment of metamodel accuracy}\label{seq:Assess_Gp}

To evaluate the accuracy of a metamodel, we use the predictivity coefficient $Q^2$: 
\begin{equation}
Q^2=1-\frac{\sum_{i=1}^{n_{\text{test}}}\left(y^{(i)}-\hat{y}^{(i)}\right)^2}{\sum_{i=1}^{n_{\text{test}}} \left( y^{(i)} - \frac{1}{n_{\text{test}}}  \sum_{i=1}^{n_{\text{test}}}  y^{(i)} \right)^2}
\end{equation}
where $(x^{(i)})_{1\leq i \leq n_{\text{test}}}$ is a test sample, $(y^{(i)})_{1\leq i \leq n_{\text{test}}}$  are the corresponding observed outputs and $(\hat{y}^{(i)})_{1\leq i \leq n_{\text{test}}}$ are the metamodel predictions. $Q^2$ corresponds to the coefficient of determination in prediction and can be computed on a test sample independent from the learning sample or by cross-validation on the learning sample. The closer to one the $Q^2$, the better the accuracy of the metamodel.
In the framework of joint Gp-modeling, $Q^2$ criterion will be used to assess the accuracy of the mean part of the joint Gp, namely $Gp_{m,\bullet}$, whether $Gp_{m,\bullet}$ is a homoscedastic ($Gp_{m,1}$) or heteroscedastic Gp ($Gp_{m,2}$). This quantitative information can be supplemented with a plot of predicted against observed values ($\hat{y}^{(i)}$ with respect to $y^{(i)}$) or a quantile-quantile plot.

To evaluate the quality of the dispersion part of a joint metamodel (denoted $Gp_{v,\bullet}$), we use the graphical tool introduced in \cite{marioo12} to assess the accuracy of the confidence intervals predicted by a Gp metamodel. For a given Gp metamodel, It consists in evaluating the proportions of observations that lie within the $\alpha$-theoretical confidence intervals which are built with the mean squared error of Gp predictions (the whole Gp structure is used and not only the conditional mean). These proportions (i.e. the observed confidence intervals) can be visualized against the $\alpha$-theoretical confidence intervals, for different values of $\alpha$.

\subsection{Application on IBLOCA test case}\label{seq:LOCA_Gp}

The joint Gp metamodel is built from the learning sample of $n = 500$: the eleven PII identified at the end of the  the screening step are considered as the explanatory variables while the sixteen others are considered as the uncontrollable parameter. Gps on mean and dispersion components are built using the sequential building process described in section \ref{methodo:sucess} where PII ordered by decreasing $R^2_{\text{HSIC}}$ are successively included in Gp. $Q^2$ coefficient of mean component $Gp_{m}$ is computed by cross validation at each iteration of the sequential building process. 
The results which are given by Table \ref{tab:Q2} show an increasing predictivity until its stabilization around 0.87, which illustrates the robustness of the Gp building process. 
The first four PII make the major contribution yielding a $Q^2$ around 0.8, the four following ones yield minor improvements (increase of 0.02 on average for each input) while the three last PII does not improve the Gp predictivity. Note that, in this application, these results remain unchanged whether we consider homoscedastic Gp ($Gp_{m,1}$) or heteroscedastic Gp ($Gp_{m,2}$), the heteroscedastic nugget effect not significantly improving the Gp predictor for the mean component. Thus, only $13\%$ of the output variability remains here not explained by $Gp_{m}$, this includes both the inaccuracy of the $Gp_{m}$ (part of $Y_m$ not fitted by Gp) and the total effect of the uncontrollable parameter, i.e. the group of non-selected inputs.


\begin{table}[!ht]
\caption{Evolution of $Gp_{m}$ metamodel predictivity during the sequential process building, for each new additional PII.}\label{tab:Q2}
\begin{tabular}{c|lllllllllll}\hline
PII &  X$_{10}$ & X$_{2}$ & X$_{12}$ & X$_{22}$ & X$_{15}$ & X$_{13}$ & X$_{9}$ & X$_{5}$ & X$_{14}$ & X$_{26}$ & X$_{27}$ \\
\hline
$Q^2$ & 0.60 & 0.64 & 0.70 & 0.79 & 0.81 & 0.83 & 0.85 & 0.85 & 0.87 & 0.87 & 0.87 \\\hline
\end{tabular}
\end{table}

For this exercise, 600 remaining CATHARE2 simulations, different from the learning sample, are also available. As they are not used to build the Gp metamodel, this set of simulations will constitutes a test sample. The $Q^2$ computed on this test sample is $Q^2 = 0.90$ for both $Gp_{m,1}$ and $Gp_{m,2}$, which is consistent with the estimations by cross-validation.\\

\begin{figure}[!ht]
      \includegraphics[width=8cm]{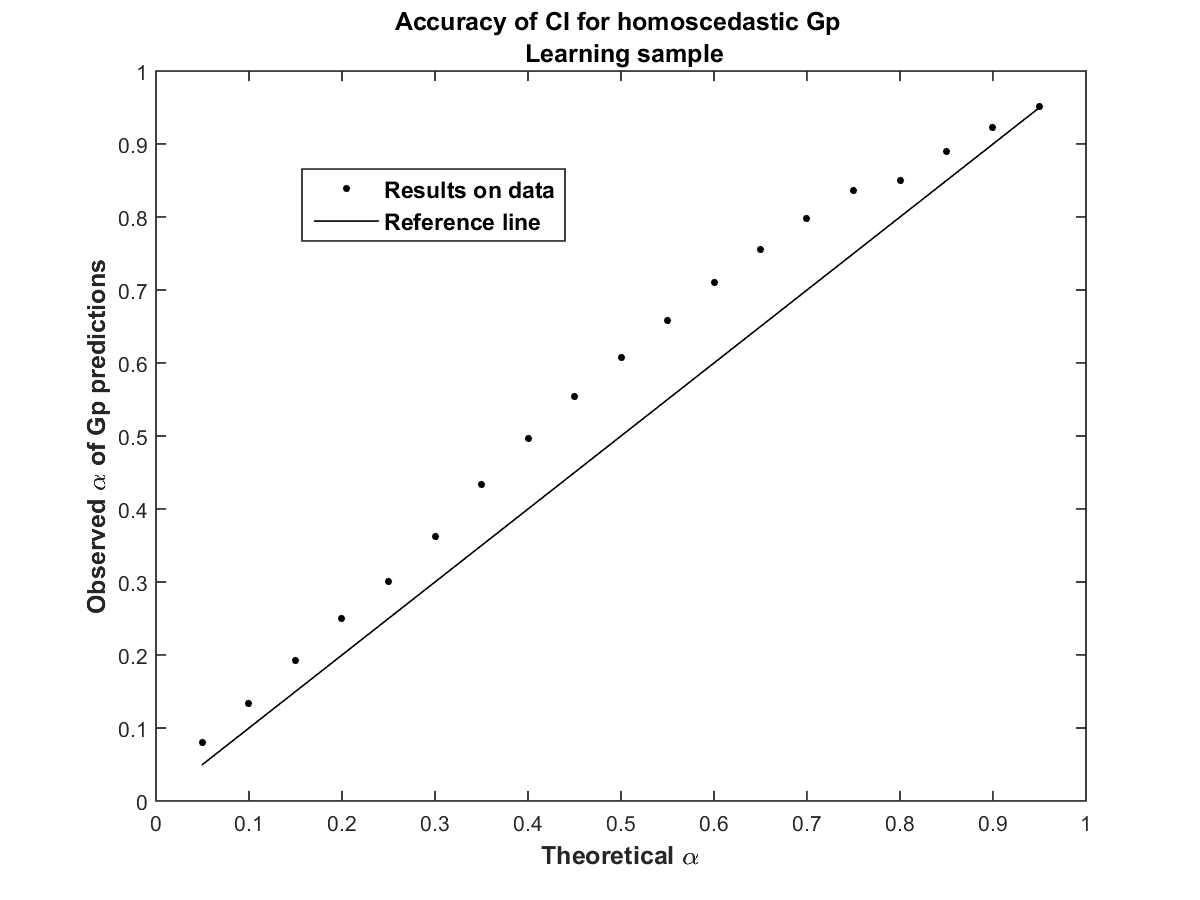}
      \includegraphics[width=8cm]{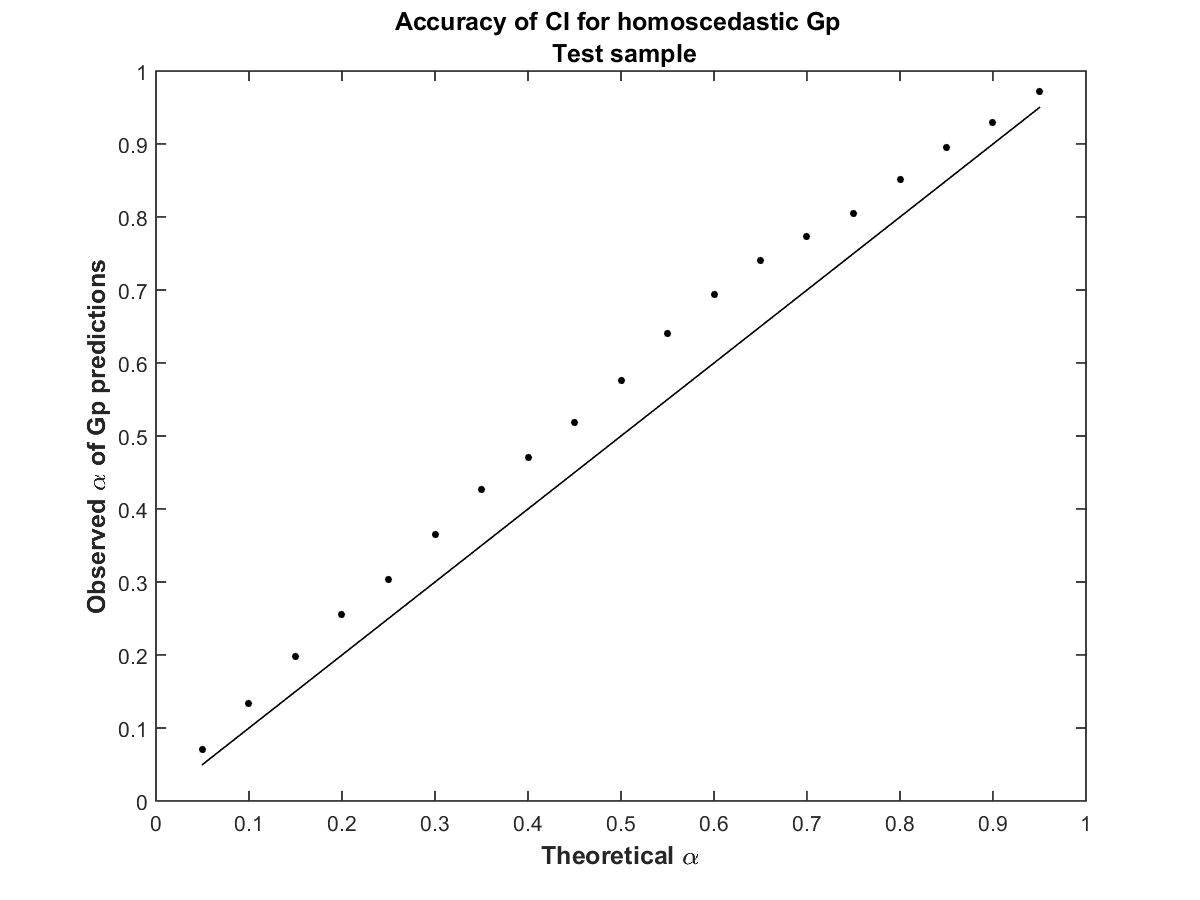}
     \\
      \includegraphics[width=8cm]{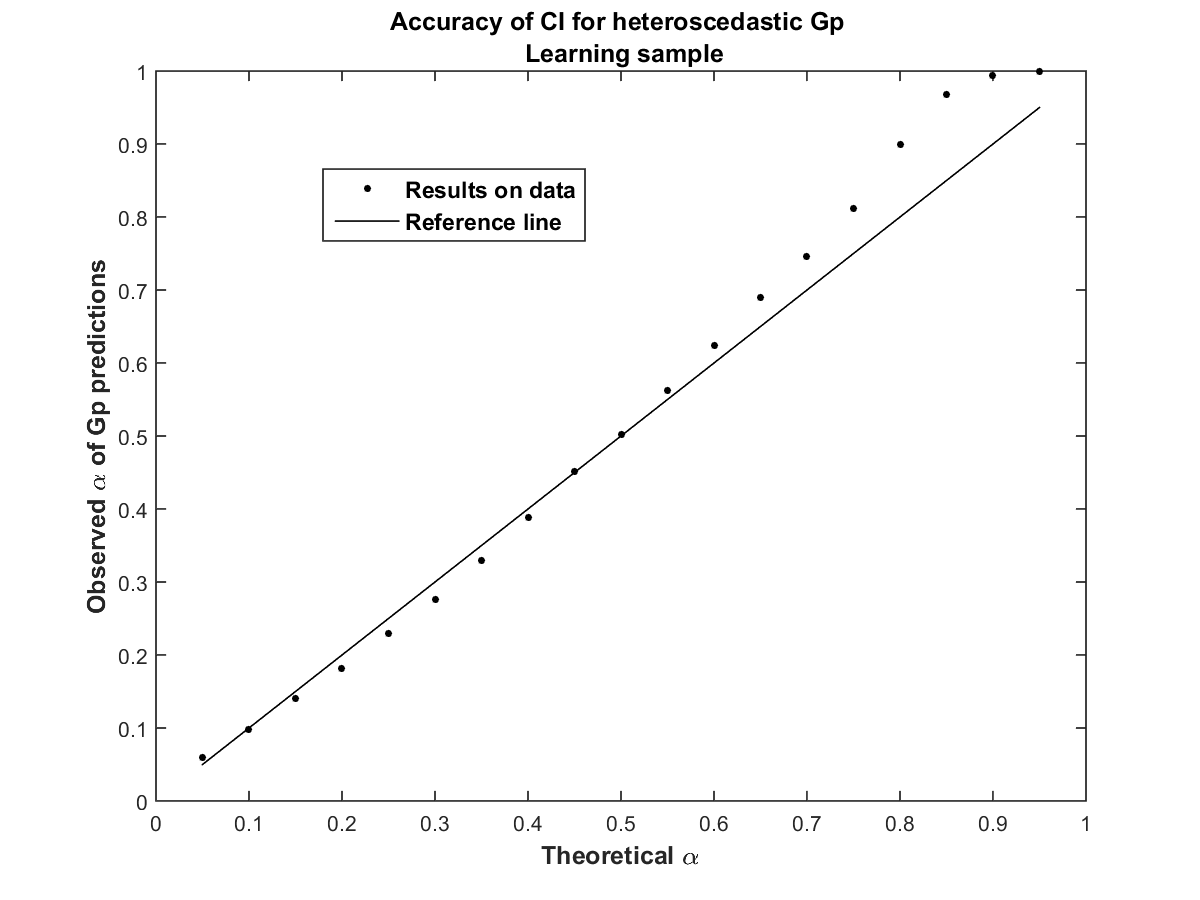}
      \includegraphics[width=8cm]{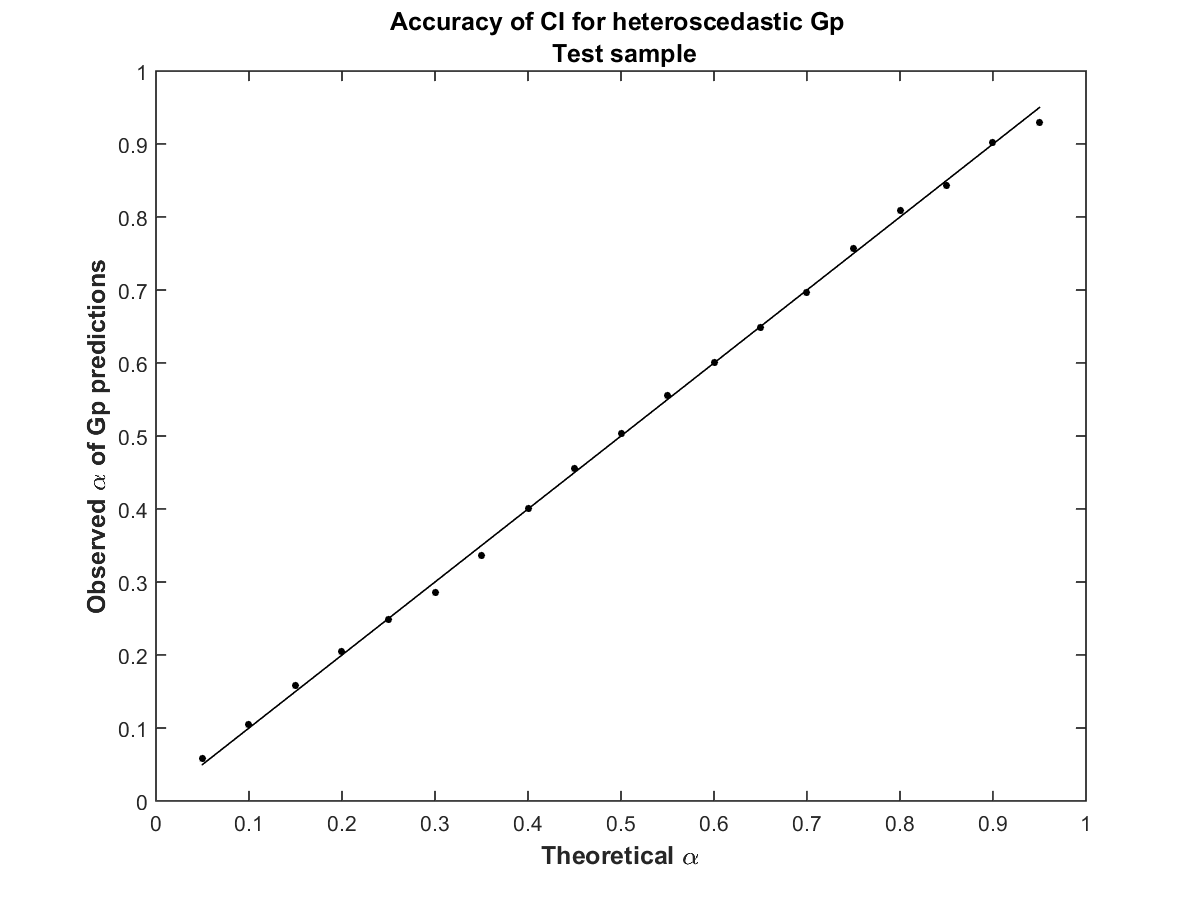}
\caption{\label{fig:ICMM}Proportion of observations that lie within the $\alpha$-confidence interval predicted by the Gp, according to the theoretical $\alpha$. Top: results for homoscedastic Gp ($Gp_{m,1}$) on the learning sample (left) and on the test sample (right).  Bottom: results for heteroscedastic Gp ($Gp_{m,2}$) on the learning sample (left) and on the test sample (right).}
\end{figure}

Now, to assess the quality of dispersion part of the joint Gp, the predicted confidence intervals are compared with the theoretical ones (cf. Section \ref{seq:Assess_Gp}). Figure \ref{fig:ICMM} gives the results obtained by $Gp_{m,1}$ and $Gp_{m,2}$ on the learning sample (by cross-validation) and the test sample, since available here. It clearly illustrates both the interest of considering a heteroscedatic nugget effect and the efficiency of using a joint Gp model to fit and predict this nugget. 
It can be seen that the joint Gp yields the most accurate confidence intervals in prediction, especially for the test sample. Indeed,  all its points are close to the theoretical $y = x$ line (a deviation is only observed for the learning sample for the highest $\alpha$), while the homoscedastic Gp tends to give too large confidence intervals. 
Thus, in this case, the heteroscedasticity hypothesis is justified and, consequently, the proposed joint Gp model is clearly more competitive than the simple one.

\section{Step 4A: Variance-based sensitivity analysis}

Sensitivity Analysis methods allow to answer the question ``How do the input parameters variations contribute, qualitatively or quantitatively, to the variation of the output?'' \cite{salrat08}. These tools can detect non-significant input parameters in a screening context, determine the most significant ones, measure their respective contributions to the output or identify an interaction between several inputs which impacts strongly the model output. From this, engineers can guide the characterization of the model by reducing the output uncertainty: for instance, they can calibrate the most influential inputs and fix the non-influential ones to nominal values. Many surveys on sensitivity analysis exist in the literature, such as \cite{kle97}, \cite{frepat02} or \cite{heljoh06}. Sensitivity analysis can be divided into two sub-domains: the local sensitivity analysis and the global sensitivity analysis. The first one studies the effects of small input perturbations around nominal values on the model output \cite{cac81a} while the second one considers the impact of the input uncertainty on the output over the whole variation domain of uncertain inputs \cite{salrat08}. 
We focus here on one of the most widely used global sensitivity indices, namely Sobol' indices which are based on output variance decomposition.


A classical approach in global sensitivity analysis is to compute the first-order and total Sobol' indices which are based on the output variance decomposition \cite{sob93,homsal96}, see Appendix \ref{sec:sobol} for mathematical details.
Sobol' indices are widely used in global sensitivity analysis because they are easy to interpret and directly usable in a dimension reduction approach.
However, their estimation (based on Monte-Carlo methods for example) requires a large number of model evaluations, which is intractable for time expensive computer codes. An classical alternative solution consists in using a metamodel to compute these indices. Note that a connection can be made between the estimation error of Sobol' indices when using a metamodel and the predictivity coefficient $Q^2$ of the metamodel. Indeed, when the $Q^2$  is estimated on a probabilized sample of the inputs (in other words when it is computed under the probability distribution of the inputs), it provides an estimation of the part of variance unexplained by the metamodel. This can be kept in mind when interpreting the Sobol' indices estimated with the metamodel.

\subsection{Sobol' indices with a joint Gp metamodel}\label{sec:GSA_joint}

In the case where a joint Gp metamodel is used to take into account an uncontrollable input $X_\varepsilon$, \cite{ioorib08} and \cite{marioo12} have shown how to deduce Sobol' indices from this joint metamodel, see Appendix \ref{sec:soboljoint} for mathematical details.

Therefore, from a joint Gp, it is only possible to estimate Sobol' indices of any subset of $\mathbf{X_{exp}}$ (equation (\ref{eqSiY})) and the total Sobol' index of $X_\varepsilon$ (equation (\ref{eqSTeps})).
The latter is interpreted as the total sensitivity index of the uncontrollable process. The individual index of $X_\varepsilon$ or any interaction index involving $X_\varepsilon$ are not directly reachable from joint Gp; their contributions in $S_{\varepsilon}^{T}$ can not be distinguished. This constitutes a limitation of this approach. However, the potential interactions between $X_\varepsilon$ and inputs of $\mathbf{X_{exp}}$ could be pointed out, considering all the primary and total effects of all the other parameters. The sensitivity analysis of $Y_d$ can also be a relevant indicator: if a subset $X_u$ of $\mathbf{X_{exp}}$ is not influential on $Y_d$, we can deduce that $S_{u\varepsilon}$ is equal to zero. Note that in practice, $\mbox{Var}(Y)$ which appears in both equations (\ref{eqSiY}) and (\ref{eqSTeps}) can be estimated directly from the learning sample (empirical estimator) or from the fitted joint Gp, using equation (\ref{decompvarcond}).

\subsection{Results on IBLOCA test case}

From the joint Gp built in section \ref{seq:LOCA_Gp}, Sobol' indices of PII are estimated from $Gp_m$ by using equation (\ref{eqSiY}), $\mbox{Var}(Y)$ being estimated with $Gp_m$ and $Gp_d$ using equation (\ref{decompvarcond}). For this, intensive Monte Carlo methods are used (see e.g. the pick-and-freeze estimator of \cite{gamjan16}): tens of thousands simulations of Gp are done. Remind that predictions of Gp are very  inexpensive (few seconds for several thousand simulations), especially with respect to the thermal-hydraulic simulator.
The obtained first Sobol' indices of PII are given by Table \ref{tab:Sobol_PII_first} and represent $85\%$ of the total variance of the output. Qualitative results of HSIC indices are confirmed and refined: X$_{10}$ remains the major influential input with $59\%$ of explained variance, followed to a lesser extend by X$_{12}$ and X$_{22}$  with for each of them $8\%$ of variance. The \textit{partial} total Sobol' indices involving only PII and derived from $Gp_m$ show that additional $4\%$ of variance is due to interaction between X$_{10}$, X$_{12}$ and X$_{22}$. The related second order Sobol' indices are estimated and the significant ones are given in Table \ref{tab:Sobol_PII_first}.  The other PII have negligible influence. 
In short,  the set of PII explain a total $89\%$ of the output variance, of which $79\%$ is only due to X$_{10}$, X$_{12}$ and X$_{22}$.

\vspace{0.3cm}

\begin{table}[!ht]
\caption{First and second Sobol' indices of PII (in \%), estimated with $Gp_{m}$ of the joint Gp metamodel.}\label{tab:Sobol_PII_first}
\begin{tabular}{c|lllllllllll}\hline
PII &  X$_{10}$ & X$_{2}$ & X$_{12}$ & X$_{22}$ & X$_{15}$ & X$_{13}$ & X$_{9}$ & X$_{5}$ & X$_{14}$ & X$_{26}$ & X$_{27}$ \\
\hline
$1^\text{st}$-order Sobol' indices & 59 & 3 & 8 & 8 & 2 & 1  & 2 & 0 & 2 & 0 & 0 \\
\hline
\end{tabular}
\vspace{0.2cm}
\\
\begin{tabular}{c|ll}
\hline
Interaction between PII &  X$_{10} \times$ X$_{22}$ & X$_{10} \times$ X$_{12}$ \\
\hline
$2^\text{nd}$-order Sobol' indices & 3 & 1 \\
\hline
\end{tabular}
\end{table}

For the physical interpretation, these results confirm those revealed in Section \ref{sec:screening}, with a rigorous quantification of inputs' importance: the interfacial friction coefficient in the horizontal part of the hot legs (X$_{10}$) is the main  contributor to the uncertainty of the PCT.
Moreover, some results have not been revealed by the HSIC-based screening analysis of Table \ref{tab:R2HSIC}.
At present, Sobol' indices clearly indicate that the interfacial friction coefficient in the SG inlet plena X$_{12}$ and the wall to liquid friction (in under-saturated break flow conditions) in the break line X$_{22}$ are more influential than the minimum stable film temperature in the core X$_{2}$.
X$_{22}$ has a significant influence on the PCT because its low values lead to higher break flow rates, resulting in a loss of the primary water mass inventory at the higher break, and thus a more significant core uncovering (then higher PCT).
For X$_{12}$, its higher values lead to a greater supply (by the vapor) of liquid possibly stored in the water plena to the rest of the primary loops (then lower PCT).
Table \ref{tab:Sobol_PII_first} also shows that there are some small interaction effects (possibly antagonistic) between X$_{10}$ and X$_{12}$, as well as between X$_{10}$ and X$_{22}$.
Let us remark that deepening this question (which is outside the scope of this paper) would be possible via plotting, from the Gp metamodel, the conditional expectations of the PCT as a function of the interaction variables.

At present, from $Gp_d$ and using equation (\ref{eqSTeps}), the total effect of the group of the sixteen non-PII inputs (i.e. $X_\varepsilon$) is estimated to $9.7\%$. 
This includes its effect alone and in interaction with the PII. To further investigate these interactions, Sobol indices of $Y_d$ are estimated and HSIC-based statistical dependence tests are applied on $Y_d$. 
They reveal that only X$_{10}$, X$_{14}$, X$_{2}$, X$_{22}$ and X$_{4}$ potentially interact with the group of non-PII inputs.
At this stage of our analysis, this result cannot be physically interpreted.

\section{Step 4B: Quantile estimation}\label{sec:quantileMM}

As already said in the introduction, once a predictive Gp metamodel has been built, it can be used to perform uncertainty propagation and in particular, estimate probabilities or, as here, quantiles. 

\subsection{Gp-based quantile estimation}
The most trivial and intuitive approach to estimate a quantile with a Gp metamodel is to apply the quantile definition to the predictor of the metamodel. 
This direct approach yields a so called \textit{plug-in} estimator. 
More precisely, with a Gp metamodel, this approach consists in using only the predictor of the Gp metamodel (i.e. the expectation conditional to the learning sample) in order to estimate the quantile. As the expectation of the Gp mean is a deterministic function of the inputs, this provides a deterministic expression of the quantile but no confidence intervals are available.
Moreover, for high (resp. low) quantiles, this methods tends to substantially underestimate (resp. overestimate) the \textit{true} quantile because the metamodel predictor is usually constructed by smoothing the computer model output values (see an analysis of this phenomenon in \cite{cangar08}).
 
To overcome this problem, \cite{oak04} has proposed to take advantage of the probabilistic-type Gp metamodel by using its entire structure: not only the mean of the conditional Gp metamodel but also its covariance structure are taken into account. 
In this \textit{full-Gp} based approach also called modular Bayesian approach, the quantile definition is therefore applied to the global Gp metamodel and leads to a random variable. 
The expectation of this random variable can be then considered as a quantile estimator. Its variance and, more generally, all its distribution can then be used as an indicator of the accuracy of the quantile estimate. Confidence intervals can be deduced, which constitutes a great advantage of this \textit{full-Gp} based approach. Moreover, the efficiency of this approach for high quantile (of the order of $95\%$) has been illustrated by \cite{rut06}. 

In practice, the estimation of quantile with the \textit{full-Gp} approach is based on stochastic simulations (conditionally to the learning sample) of the Gp metamodel, by using the standard method of conditional simulations \cite{chidel99}. 
We recall just that, to generate conditional Gaussian simulations on a given interval, a discretization of the interval is first considered, then the vector of the conditional expectation and the matrix of the conditional covariance on the discretized interval are required.
Note that \cite{legcan14} uses this approach with Gp conditional simulations for the estimation of Sobol' indices and their associated confidence intervals.

In this paper, from our joint Gp model, we will compare the \textit{full-Gp} approach applied to either the homoscedastic Gp or the heteroscedastic Gp and will proceed as follows: 
\begin{itemize}
\item For the homoscedastic Gp, the standard technique of conditional simulations is applied to $Gp_{m,1}$ (built to estimate $Y_m$, with a constant nugget effect).
\item For the heteroscedastic Gp, we propose a new technique to simulate the conditional Gp trajectories: 
\begin{enumerate}
\item The heteroscedastic Gp built for $Y_m$ (namely $Gp_{m,2}$) provides the conditional expectation vector and a preliminary conditional covariance matrix.
\item The Gp built for $Y_d$ (namely $Gp_{d,2}$) is predicted and provides the heteroscedastic nugget effect which is added to the diagonal of the conditional covariance matrix of the previous step.
\item Conditional simulations are then done using the standard method \cite{chidel99}.
\end{enumerate}
\end{itemize}

\subsection{Results on IBLOCA test case}

In this section, we focus on the estimation of the $95\%$-quantile of the PCT for the IBLOCA application.
From the learning sample of size $n=500$ and the joint Gp, we compare here the following approaches to estimate the PCT quantile:
\begin{itemize}
\item The classical empirical quantile estimator, denoted $ \hat{q}_{95}^{\text{emp}}$. A bootstrap method (see for example \cite{ziodim08}) makes it possible to obtain in addition a $90\%$-confidence interval for this empirical quantile.
\item The \textit{plug-in} estimators from the homoscedastic or the heteroscedastic Gp, denoted $\hat{q}_{95}^{Gp_{m,1}\text{-pi}}$ and $\hat{q}_{95}^{Gp_{m,2}\text{-pi}}$. No confidence intervals are obtained using this estimation method.
\item The \textit{full-Gp} estimators from the homoscedastic or the heteroscedastic Gp, denoted $\hat{q}_{95}^{Gp_{m,1}\text{-full}}$ and $\hat{q}_{95}^{Gp_{m,2}\text{-full}}$ respectively. As explained in the previous section, confidence intervals can be deduced with this \textit{full-Gp} approach.
\end{itemize}

Table \ref{tab:results} synthesizes all the results obtained for the PCT $95\%$-quantile with these different approaches given above. In addition, $90\%$-confidence intervals are given when they are available.

As explained in Section \ref{seq:LOCA_Gp}, we also have 600 other CATHARE2 simulations, for a total of 1100 PCT simulations (learning sample plus test sample). A reference value of the $95\%$-quantile is obtained from this full sample with the classical empirical quantile estimator:
\[ \hat{q}_{95}^{\text{ref}} = 742.28\text{ }^{\circ}\text{C} \;.\]
The empirical estimator based on the learning sample is imprecise but its bootstrap-based confidence interval is consistent with the reference value. As expected, the \textit{plug-in} approaches provide poor estimations of the quantile, which is here strongly underestimated. 
The \textit{full-Gp} approach based on the homoscedastic assumption overestimates the quantile; this is consistent with the analysis of Gp confidence intervals in Figure \ref{fig:ICMM} (too large confidence intervals provided by homoscedastic Gp).
Finally, the best result is obtained with the conditional simulation method based on the heteroscedastic Gp metamodel, built with the joint Gp method. 
This approach yields a more accurate prediction than the usual homoscedastic Gp and outperforms the empirical estimator in terms of confidence interval. Once again, the heteroscedasticity hypothesis is clearly relevant in this case.

\begin{table}[!ht]
\caption{Results for the $95\%$-quantile estimates of the PCT (in $\text{ }^{\circ}\text{C}$) with its $90\%$-confidence interval (CI) when available.}\label{tab:results}
\begin{tabular}{c|c|c|cc|cc}\hline
\rule[-0.2cm]{0cm}{0.2cm} & \multirow{2}{*}{Reference\tablefootnote{\samepage The reference method uses the full sample of size $n=1100$.}}  & \multirow{2}{*}{Empirical\tablefootnote{\label{note1}Empirical and Gp-based methods are applied from the learning sample of size $n=500$.}} & \multicolumn{2}{c|}{\textit{Plug-in}\footref{note1}$ ^{,}$\tablefootnote{\label{note2}The \textit{plug-in} and \textit{full-Gp} estimators are based on either the homoscedastic or heteroscedastic Gp metamodel.}
} & \multicolumn{2}{c}{\textit{Full-Gp}\footref{note1}$ ^{,}$\footref{note2}
 (conditional simulations)}  \\
\rule[-0.2cm]{0cm}{0.2cm} &  &  & homo-Gp & hetero-Gp & homo-Gp & hetero-Gp\\
\rule[-0.2cm]{0cm}{0.2cm}  & $\hat{q}_{95}^{\text{ref}}$ & $\hat{q}_{95}^{\text{emp}}$ & $\hat{q}_{95}^{Gp_{m,1}\text{-pi}}$ & $\hat{q}_{95}^{Gp_{m,2}\text{-pi}}$ & $\hat{q}_{95}^{Gp_{m,1}\text{-full}}$ & $\hat{q}_{95}^{Gp_{m,2}\text{-full}}$ \\
\hline\hline
\rule[-0.2cm]{0cm}{0.2cm}Mean  & $\mathbf{742.28}$ & $746.80$ & $736.26$ & $735.83$ & $747.11$ & $\mathbf{741.46}$ \\
\rule[-0.2cm]{0cm}{0.2cm}$90\%$-CI & --- & $[ 736.7 ; 747.41 ]$ & --- & --- & $[ 742.93 ; 751.32 ]$ & $\mathbf{[ 738.76 ; 744.17 ]}$
\end{tabular}\end{table}

\section{Conclusion}

In the framework of the estimation of safety margins in nuclear accident analysis, it is essential to quantitatively assess the uncertainties tainting the results of Best-estimate codes. In this context, this paper has been focused on an advanced statistical methodology for Best Estimate Plus Uncertainty (BEPU) analysis, illustrated by a high dimensional thermal-hydraulic test case simulating accidental scenario in a Pressurized Water Reactor (IBLOCA test case). 
Some statistical analyses such as the estimation of high-level quantiles or quantitative sensitivity analysis (e.g., estimation of Sobol' indices based on variance decomposition) may call in practice for several thousand of code simulations. Complex computer codes, as the ones used in thermal-hydraulic accident scenario simulations, are often too cpu-time expensive to be directly used to perform these studies.

To cope with this limitation, we propose a methodology mainly based on a predictive joint Gp metamodel, built with an efficient sequential algorithm.
First, an initial screening step based on advanced dependence measures and associated statistical tests enabled to identify a group of significant inputs, allowing a reduction of the dimension. 
The efforts of optimization when fitting the metamodel can then be concentrated on the main influential inputs.
The robustness of metamodeling is thus increased. 
Moreover, thanks to the joint metamodel approach, the non-selected inputs are not completely removed: the residual uncertainty due to dimension reduction is integrated in the metamodel and the global influence of non-selected inputs is so controlled. From this joint Gp metamodel, accurate uncertainty propagation and quantitative sensitivity analysis, not feasible with the numerical model because of its computational cost, become accessible. Thus, the uncertainties of model inputs are propagated inside the joint Gp to estimate Sobol' indices, failure probabilities and/or quantiles, without additional code simulations. 

Thus, on the IBLOCA application, a predictive Gp metamodel is built with only a few hundred of code simulations (500 code simulations for 27 uncertain inputs). From this joint Gp, a quantitative sensitivity analysis based on variance decomposition is performed without additional code simulation: Sobol' indices are computed and reveal that the output is mainly explained by four uncertain inputs. One of them, namely the interfacial friction coefficient in the hot legs, is strongly influential with around $60\%$ of output variance explained, the three others being of minor influence.
The quite less influence of all the other inputs is also confirmed. Note that a direct and accurate computation of Sobol' indices with the thermal-hydraulic code would have required tens of thousands of code simulations.

The physical interpretation of the results obtained with the screening step and the variance-based sensitivity analysis step are known to be useful for modelers and engineers.
This study has demonstrated this once again, in the particular case of an IBLOCA safety analysis, by revealing some physical effects on the PCT of influential inputs which cannot be understood without a global statistical approach (e.g. the threshold effect due to the interfacial friction coefficient in the horizontal part of the hot legs).
\cite{geivac18} has also underlined the importance of sensitivity analysis in a validation methodology in order to identify relevant physical model uncertainties on which safety engineers must focus their efforts.
Counter-intuitive effects are also present during an IBLOCA transient and only a global understanding of physical phenomena can help.
As a perspective of the present work, extending the screening and sensitivity analysis tools to a joint analysis of several relevant output variables of interest (as the PCT time, the primary pressure and the core differential pressure) would be essential.

In the IBLOCA test case, we are particularly interested by the estimation of the $95\%$-quantile of the model output temperature.
In nuclear safety, as in other engineering domains, methods of conservative computation of quantiles \cite{nutwal04,hesuhl15} have been largely studied. 
We have shown in the present work how to use and simulate the joint Gp metamodel to reach the same objective: the uncertainty of the influential inputs are directly and accurately propagated through the mean component of the joint metamodel while a confidence bound is derived from the dispersion component in order to take into account the residual uncertainty of the other inputs. Results on the IBLOCA test case show that joint Gp provides a more accurate estimation of the $95\%$-quantile than the empirical approach, at equal calculation budget. Besides, this estimation is very close from the reference value obtained with 600 additional code simulations. Furthermore, the interest of the heteroscedastic approach in joint Gp is also emphasized: the estimated quantile and the associated confidence interval are much better than those of the homoscedastic approach. 

As a future application of modern statistical methods on IBLOCA safety issues, one should mention the use of Gp metamodels to identify penalizing thermal-hydraulic transients, with respect to some particular scenario inputs.
As in the present work, strong difficulties are raised by the cpu time cost of the computer code and the large number of uncertain (and uncontrollable) inputs.

\section*{Acknowledgments}

We are grateful to Henri Geiser and Thibault Delage who performed the computations of the CATHARE2 code.

\pagebreak

\begin{thebibliography}{10}
\newcommand{\enquote}[1]{``#1''}
\providecommand{\url}[1]{\texttt{#1}}
\providecommand{\urlprefix}{URL }
\expandafter\ifx\csname urlstyle\endcsname\relax
  \providecommand{\doi}[1]{doi:\discretionary{}{}{}#1}\else
  \providecommand{\doi}{doi:\discretionary{}{}{}\begingroup
  \urlstyle{rm}\Url}\fi

\bibitem{poumod09}
\textsc{M.~{Pourgol-Mohamad}}, \textsc{M.~Modarres}, and \textsc{A.~Mosleh},
  \enquote{Integrated methodology for thermal-hydraulic code uncertainty
  analysis with application,} \emph{Nuclear Technology}, \textbf{165}, 333
  (2009).

\bibitem{bucpet10}
\textsc{A.~Bucalossi}, \textsc{A.~Petruzzi}, \textsc{M.~Kristof}, and
  \textsc{F.~{D'Auria}}, \enquote{Comparison between
  best-estimate-plus-uncertainty methods and conservative tools for nuclear
  power plant licensing,} \emph{Nuclear Technology}, \textbf{172}, 29 (2010).

\bibitem{promav07}
\textsc{A.~Prosek} and \textsc{B.~Mavko}, \enquote{The state-of-the-art theory
  and applications of best-estimate plus uncertainty methods,} \emph{Nuclear
  Technology}, \textbf{158}, 69 (2007).

\bibitem{wil13}
\textsc{G.~Wilson}, \enquote{Historical insights in the development of {B}est
  {e}stimate {P}lus {U}ncertainty safety analysis,} \emph{Annals of Nuclear
  Energy}, \textbf{52}, 2 (2013).

\bibitem{sansan18}
\textsc{F.~{Sanchez-Saez}}, \textsc{A.~S\`anchez}, \textsc{J.~Villanueva},
  \textsc{S.~Carlos}, and \textsc{S.~Martorell}, \enquote{Uncertainty analysis
  of large break loss of coolant accident in a pressurized water reactor using
  non-parametric methods,} \emph{Reliability Engineering and System Safety},
  \textbf{174}, 19 (2018).

\bibitem{nutwal04}
\textsc{W.~Nutt} and \textsc{G.~Wallis}, \enquote{Evaluation of nuclear safety
  from the outputs of computer codes in the presence of uncertainties,}
  \emph{Reliability Engineering and System Safety}, \textbf{83}, 57 (2004).

\bibitem{wal07}
\textsc{G.~Wallis}, \enquote{Uncertainties and probabilities in nuclear reactor
  regulation,} \emph{Nuclear Engineering and Design}, \textbf{237}, 1586
  (2004).

\bibitem{decbaz08}
\textsc{A.~{de Cr\'ecy}}, \textsc{P.~Bazin}, \textsc{H.~Glaeser},
  \textsc{T.~Skorek}, \textsc{J.~Joucla}, \textsc{P.~Probst},
  \textsc{K.~Fujioka}, \textsc{B.~Chung}, \textsc{D.~Oh}, \textsc{M.~Kyncl},
  \textsc{R.~Pernica}, \textsc{J.~Macek}, \textsc{R.~Meca}, \textsc{R.~Macian},
  \textsc{F.~D'Auria}, \textsc{A.~Petruzzi}, \textsc{L.~Batet},
  \textsc{M.~Perez}, and \textsc{F.~Reventos}, \enquote{Uncertainty and
  sensitivity analysis of the {LOFT L2-5} test: {R}esults of the {BEMUSE}
  programme,} \emph{Nuclear Engineering and Design}, \textbf{12}, 3561 (2008).

\bibitem{petdau08}
\textsc{A.~Petruzzi} and \textsc{F.~{D'Auria}}, \enquote{Approaches, relevant
  topics, and internal method for uncertainty evaluation in predictions of
  thermal-hydraulic system codes,} \emph{Science and Technology of Nuclear
  Installations}, \textbf{2008}, \emph{Article ID 325071, 17 pages,
  DOI:10.1155/2008/325071} (2008).

\bibitem{marnut11}
\textsc{R.~Martin} and \textsc{W.~Nutt}, \enquote{Perspectives on the
  application of order-statistics in best-estimate plus uncertainty nuclear
  safety analysis,} \emph{Nuclear Engineering and Design}, \textbf{241}, 274
  (2011).

\bibitem{derdev08}
\textsc{E.~{de Rocquigny}}, \textsc{N.~Devictor}, and \textsc{S.~Tarantola}
  (Editors), \emph{Uncertainty in industrial practice}, Wiley (2008).

\bibitem{fanli06}
\textsc{K.-T. Fang}, \textsc{R.~Li}, and \textsc{A.~Sudjianto}, \emph{Design
  and modeling for computer experiments}, Chapman \& Hall/CRC (2006).

\bibitem{forsob08}
\textsc{A.~Forrester}, \textsc{A.~Sobester}, and \textsc{A.~Keane} (Editors),
  \emph{Engineering design via surrogate modelling: a practical guide}, Wiley
  (2008).

\bibitem{cangar08}
\textsc{C.~Cannamela}, \textsc{J.~Garnier}, and \textsc{B.~Iooss},
  \enquote{Controlled stratification for quantile estimation,} \emph{Annals of
  Apllied Statistics}, \textbf{2}, 1554 (2008).

\bibitem{ziodim09}
\textsc{E.~Zio}, \textsc{F.~{Di Maio}}, \textsc{S.~Martorell}, and
  \textsc{Y.~Nebot}, \enquote{Neural network and Order Statistics for
  Quantifying nuclear power plant Safety Margins,} \textsc{S.~Martorell},
  \textsc{C.~G. Soares}, and \textsc{J.~Barnett} (Editors), \emph{Safety,
  reliability and risk analysis - Proceedings of the ESREL 2008 Conference},
  2873--2881, CRC Press (2009).

\bibitem{lorzan11}
\textsc{G.~Lorenzo}, \textsc{P.~Zanocco}, \textsc{M.~Gim\'enez},
  \textsc{M.Marqu\`es}, \textsc{B.~Iooss}, \textsc{R.~{Bolado-Lavin}},
  \textsc{F.~Pierro}, \textsc{G.~Galassi}, \textsc{F.~{D’Auria}}, and
  \textsc{L.~Burgazzi}, \enquote{Assessment of an isolation condenser of an
  integral reactor in view of uncertainties in engineering parameters,}
  \emph{Science and Technology of Nuclear Installations}, \textbf{2011},
  \emph{Article ID 827354, 9 pages, DOI: 10.1155/2011/827354} (2011).

\bibitem{ghahig17}
\textsc{R.~Ghanem}, \textsc{D.~Higdon}, and \textsc{H.~Owhadi} (Editors),
  \emph{Springer Handbook on {Uncertainty Quantification}}, Springer (2017).

\bibitem{sanwil03}
\textsc{T.~Santner}, \textsc{B.~Williams}, and \textsc{W.~Notz}, \emph{The
  design and analysis of computer experiments}, Springer (2003).

\bibitem{muerou12}
\textsc{T.~Muehlenstaedt}, \textsc{O.~Roustant}, \textsc{L.~Carraro}, and
  \textsc{S.~Kuhnt}, \enquote{Data-driven {K}riging models based on
  {FANOVA}-decomposition,} \emph{{Statistics \& Computing}}, \textbf{22}, 723
  (2012).

\bibitem{durgin13}
\textsc{N.~Durrande}, \textsc{D.~G. O.}, \textsc{Roustant}, and
  \textsc{L.~Carraro}, \enquote{{ANOVA kernels and RKHS} of zero mean functions
  for model-based sensitivity analysis,} \emph{Journal of Multivariate
  Analysis}, \textbf{155}, 57 (2013).

\bibitem{welbuc92}
\textsc{W.~Welch}, \textsc{R.~Buck}, \textsc{J.~Sacks}, \textsc{H.~Wynn},
  \textsc{T.~Mitchell}, and \textsc{M.~Morris}, \enquote{Screening, predicting,
  and computer experiments,} \emph{Technometrics}, \textbf{34}, \emph{1}, 15
  (1992).

\bibitem{marioo08}
\textsc{A.~Marrel}, \textsc{B.~Iooss}, \textsc{F.~{Van Dorpe}}, and
  \textsc{E.~Volkova}, \enquote{An efficient methodology for modeling complex
  computer codes with {G}aussian processes,} \emph{Computational Statistics and
  Data Analysis}, \textbf{52}, 4731 (2008).

\bibitem{woolew17}
\textsc{D.~Woods} and \textsc{S.~Lewis}, \enquote{Design of experiments for
  screening,} \textsc{R.~Ghanem}, \textsc{D.~Higdon}, and \textsc{H.~Owhadi}
  (Editors), \emph{Springer Handbook on {Uncertainty Quantification}},
  1143--1185, Springer (2017).

\bibitem{dav15}
\textsc{S.~{Da Veiga}}, \enquote{Global sensitivity analysis with dependence
  measures,} \emph{Journal of Statistical Computation and Simulation},
  \textbf{85}, 1283 (2015).

\bibitem{delmar16b}
\textsc{M.~{De Lozzo}} and \textsc{A.~Marrel}, \enquote{New improvements in the
  use of dependence measures for sensitivity analysis and screening,}
  \emph{Journal of Statistical Computation and Simulation}, \textbf{86}, 3038
  (2016).

\bibitem{ioomar17}
\textsc{B.~Iooss} and \textsc{A.~Marrel}, \enquote{An efficient methodology for
  the analysis and metamodeling of computer experiments with large number of
  inputs,} \emph{Proceedings of UNCECOMP 2017 Conference}, Rhodes Island,
  Greece (2017).

\bibitem{marioo12}
\textsc{A.~Marrel}, \textsc{B.~Iooss}, \textsc{S.~{Da~Veiga}}, and
  \textsc{M.~Ribatet}, \enquote{Global sensitivity analysis of stochastic
  computer models with joint metamodels,} \emph{Statistics and Computing},
  \textbf{22}, 833 (2012).

\bibitem{zabdej98}
\textsc{I.~Zabalza}, \textsc{J.~Dejean}, and \textsc{D.~Collombier},
  \enquote{Prediction and density estimation of a horizontal well productivity
  index using generalized linear models,} \emph{ECMOR VI, Peebles} (1998).

\bibitem{mazvac16}
\textsc{P.~Mazgaj}, \textsc{J.-L. Vacher}, and \textsc{S.~Carnevali},
  \enquote{Comparison of {CATHARE} results with the experimental results of
  cold leg intermediate break {LOCA} obtained during {ROSA-2/LSTF} test 7,}
  \emph{EPJ Nuclear Sciences \& Technology}, \textbf{2}, \emph{1} (2016).

\bibitem{geivac17}
\textsc{H.~Geiser}, \textsc{J.-L. Vacher}, and \textsc{P.~Rubiolo},
  \enquote{The use of integral effect tests for the justification of new
  evaluation models based on the {BEPU} approach,} \emph{Proceedings of
  NURETH-17}, Xi'an, Shaanxi, China (2017).

\bibitem{mckbec79}
\textsc{M.~{McKay}}, \textsc{R.~Beckman}, and \textsc{W.~Conover}, \enquote{A
  comparison of three methods for selecting values of input variables in the
  analysis of output from a computer code,} \emph{Technometrics}, \textbf{21},
  239 (1979).

\bibitem{damcou13}
\textsc{G.~Damblin}, \textsc{M.~Couplet}, and \textsc{B.~Iooss},
  \enquote{Numerical studies of space filling designs: {O}ptimization of
  {L}atin hypercube samples and subprojection properties,} \emph{Journal of
  Simulation}, \textbf{7}, 276 (2013).

\bibitem{josgul15}
\textsc{V.~Joseph}, \textsc{E.~Gul}, and \textsc{S.~Ba}, \enquote{Maximum
  projection designs for computer experiments,} \emph{Biometrika},
  \textbf{102}, 371 (2015).

\bibitem{jinche05}
\textsc{R.~Jin}, \textsc{W.~Chen}, and \textsc{A.~Sudjianto}, \enquote{An
  efficient algorithm for constructing optimal design of computer experiments,}
  \emph{Journal of Statistical Planning and Inference}, \textbf{134}, 268
  (2005).

\bibitem{loesac09}
\textsc{J.~Loeppky}, \textsc{J.~Sacks}, and \textsc{W.~Welch},
  \enquote{Choosing the sample size of a computer experiment: {A} practical
  guide,} \emph{Technometrics}, \textbf{51}, 366 (2009).

\bibitem{par62}
\textsc{E.~Parzen}, \enquote{On Estimation of a Probability Density Function
  and Mode,} \emph{The Annals of Mathematical Statistics}, \textbf{33},
  \emph{3}, 1065 (1962).

\bibitem{kucioo17}
\textsc{S.~Kucherenko} and \textsc{B.~Iooss}, \enquote{Derivative-Based Global
  Sensitivity Measures,} \textsc{R.~Ghanem}, \textsc{D.~Higdon}, and
  \textsc{H.~Owhadi} (Editors), \emph{Springer Handbook on {Uncertainty
  Quantification}}, 1241--1263, Springer (2017).

\bibitem{roubar17}
\textsc{O.~Roustant}, \textsc{F.~Barthe}, and \textsc{B.~Iooss},
  \enquote{Poincar\'e inequalities on intervals - application to sensitivity
  analysis,} \emph{Electronic Journal of Statistics}, \textbf{2}, 3081 (2017).

\bibitem{ioolem15}
\textsc{B.~Iooss} and \textsc{P.~Lema\^{\i}tre}, \enquote{A review on global
  sensitivity analysis methods,} \textsc{C.~Meloni} and \textsc{G.~Dellino}
  (Editors), \emph{Uncertainty management in Simulation-Optimization of Complex
  Systems: Algorithms and Applications}, 101--122, Springer (2015).

\bibitem{grebou05}
\textsc{G.~Gretton}, \textsc{O.~Bousquet}, \textsc{A.~Smola}, and
  \textsc{B.~Sch\"olkopf}, \enquote{Measuring statistical dependence with
  {Hilbert-Schmidt} norms,} \emph{Proceedings Algorithmic Learning Theory},
  63--77, Springer-Verlag (2005).

\bibitem{sacwel89}
\textsc{J.~Sacks}, \textsc{W.~Welch}, \textsc{T.~Mitchell}, and
  \textsc{H.~Wynn}, \enquote{Design and analysis of computer experiments,}
  \emph{Statistical Science}, \textbf{4}, 409 (1989).

\bibitem{raswil06}
\textsc{C.~Rasmussen} and \textsc{C.~Williams}, \emph{Gaussian processes for
  machine learning}, MIT Press (2006).

\bibitem{salrat08}
\textsc{A.~Saltelli}, \textsc{M.~Ratto}, \textsc{T.~Andres},
  \textsc{F.~Campolongo}, \textsc{J.~Cariboni}, \textsc{D.~Gatelli},
  \textsc{M.~Salsana}, and \textsc{S.~Tarantola}, \emph{Global sensitivity
  analysis - The primer}, Wiley (2008).

\bibitem{kle97}
\textsc{J.~Kleijnen}, \enquote{Sensitivity analysis and related analyses: a
  review of some statistical techniques,} \emph{Journal of Statistical
  Computation and Simulation}, \textbf{57}, 111 (1997).

\bibitem{frepat02}
\textsc{H.~Frey} and \textsc{S.~Patil}, \enquote{Identification and review of
  sensitivity analysis methods,} \emph{Risk Analysis}, \textbf{22}, 553 (2002).

\bibitem{heljoh06}
\textsc{J.~Helton}, \textsc{J.~Johnson}, \textsc{C.~Salaberry}, and
  \textsc{C.~Storlie}, \enquote{Survey of sampling-based methods for
  uncertainty and sensitivity analysis,} \emph{Reliability Engineering and
  System Safety}, \textbf{91}, 1175 (2006).

\bibitem{cac81a}
\textsc{D.~Cacuci}, \enquote{Sensitivity theory for nonlinear systems. {I.
  N}onlinear functional analysis approach,} \emph{Journal of Mathematical
  Physics}, \textbf{22}, 2794 (1981).

\bibitem{sob93}
\textsc{I.~Sobol}, \enquote{Sensitivity estimates for non linear mathematical
  models,} \emph{Mathematical Modelling and Computational Experiments},
  \textbf{1}, 407 (1993).

\bibitem{homsal96}
\textsc{T.~Homma} and \textsc{A.~Saltelli}, \enquote{Importance measures in
  global sensitivity analysis of non linear models,} \emph{Reliability
  Engineering and System Safety}, \textbf{52}, 1 (1996).

\bibitem{ioorib08}
\textsc{B.~Iooss} and \textsc{M.~Ribatet}, \enquote{Global sensitivity analysis
  of computer models with functional inputs,} \emph{Reliability Engineering and
  System Safety}, \textbf{94}, 1194 (2009).

\bibitem{gamjan16}
\textsc{F.~Gamboa}, \textsc{A.~Janon}, \textsc{T.~Klein}, \textsc{A.~Lagnoux},
  and \textsc{C.~Prieur}, \enquote{Statistical inference for Sobol pick freeze
  {M}onte {C}arlo methods,} \emph{Statistics}, \textbf{50}, 881 (2016).

\bibitem{oak04}
\textsc{J.~Oakley}, \enquote{Estimating percentiles of uncertain computer code
  outputs,} \emph{Applied Statistics}, \textbf{53}, 83 (2004).

\bibitem{rut06}
\textsc{B.~Rutherford}, \enquote{A response-modeling alternative to surrogate
  models for support in computational analyses,} \emph{Reliability Engineering
  and System Safety}, \textbf{91}, 1322 (2006).

\bibitem{chidel99}
\textsc{J.-P. Chil\`es} and \textsc{P.~Delfiner}, \emph{Geostatistics: Modeling
  spatial uncertainty}, Wiley, New-York (1999).

\bibitem{legcan14}
\textsc{L.~{Le Gratiet}}, \textsc{C.~Cannamela}, and \textsc{B.~Iooss},
  \enquote{A {B}ayesian approach for global sensitivity analysis of
  (multifidelity) computer codes,} \emph{SIAM/ASA Journal on Uncertainty
  Quantification}, \textbf{2}, 336 (2014).

\bibitem{ziodim08}
\textsc{E.~Zio} and \textsc{F.~{Di Maio}}, \enquote{Bootstrap and Order
  Statistics for Quantifying Thermal-Hydraulic Code Uncertainties in the
  Estimation of Safety Margins,} \emph{Science and Technology of Nuclear
  Installations}, \textbf{2008}, \emph{Article ID 340164, 9 pages,
  DOI:10.1155/2008/340164} (2008).

\bibitem{geivac18}
\textsc{H.~Geiser}, \textsc{J.-L. Vacher}, and \textsc{P.~Rubiolo},
  \enquote{Sensitivity analysis applied to {LOCA} integral effect tests for the
  justification of the {BEPU} approach,} \emph{ANS Best Estimate Plus
  Uncertainty International Conference (BEPU 2018)}, Lucca, Italia (2018).

\bibitem{hesuhl15}
\textsc{J.~Hessling} and \textsc{J.~Uhlmann}, \enquote{Robustness of {W}ilks'
  conservative estimate of confidence intervals,} \emph{International Journal
  for Uncertainty Quantification}, \textbf{5}, 569 (2015).

\bibitem{hoe48}
\textsc{W.~Hoeffding}, \enquote{A class of statistics with asymptotically
  normal distributions,} \emph{Annals of Mathematical Statistics}, \textbf{19},
  293 (1948).

\end{thebibliography}


\pagebreak
\appendix

\section{Appendix: HSIC dependence measures}\label{sec:HSIC}

If we consider two reproducing kernel Hilbert spaces $\mathcal{F}_k$ and $\mathcal{G}$ of functions $\mathcal{X}_k \rightarrow \mathbb{R}$ and $\mathcal{Y} \rightarrow \mathbb{R}$ respectively, the crossed-covariance $C_{X_k, Y}$ operator associated to the joint probabilistic distribution of $\left(X_k,Y\right)$ is the linear operator defined for every $f_{X_k} \in \mathcal{F}_k$ and $g_Y \in \mathcal{G}$ by: 
\begin{equation}
\langle f_{X_k},C_{X_k, Y} g_{Y} \rangle_{\mathcal{F}_k}=\text{Cov}\left(f_{X_k},g_{Y}\right).
\end{equation}
$C_{X_k, Y}$ generalizes the covariance matrix by representing higher order correlations between $X_k$ and $Y$ through nonlinear kernels. The HSIC criterion is then defined by the Hilbert-Schmidt norm of the
cross-covariance operator:
\begin{equation}
\text{HSIC}(X_k,Y)_{\mathcal{F}_k,\mathcal{G}} = \|C_k\|_{HS}^2.
\end{equation}
From this, \cite{dav15} introduces a normalized version of the HSIC which provides a sensitivity index of $X_k$, lying in $[0,1]$:
\begin{equation}\label{eq:HSICRk}
 R^2_{\text{HSIC},k}=\frac{\text{HSIC}(X_k,Y)}{\sqrt{\text{HSIC}(X_k,X_k)\text{HSIC}(Y,Y)}}.
\end{equation}
The closer to one the $R^2_{\text{HSIC},k}$, the stronger the dependence between $X_k$ and $Y$. 
In practice, \cite{grebou05} propose a Monte Carlo estimator of $\text{HSIC}(X_k,Y)$ and a plug-in estimator can be deduced for $R^2_{\text{HSIC},k}$. Note that Gaussian kernel functions with empirical estimations of the variance parameter are used in our application  (see \cite{grebou05} for more details).

\section{Appendix: Sobol' indices}\label{sec:sobol}

If $X=(X_1,\ldots,X_d)$ with independence between the variables $X_1,\ldots,X_d$ and if $\E[g^2(X)]<+\infty$, we can apply the Hoeffding decomposition to the random variable $g(X)$ \cite{hoe48}:
\begin{eqnarray}\label{eq:hoeffding}
g(X) &=&\sum_{u\subset\{1,\ldots,d\}}g_u(X_u)
\end{eqnarray}
where $g_{\emptyset}=\E[g(X)]$, $g_i(X_i)=\E[g(X)|X_i]-g_{\emptyset}$ and $g_u(X_u)=\E[g(X)|X_u]-\sum_{v \subsetneq u}g_v(X_v)$, with $X_u=(X_i)_{i\in u}$, for all $u\subset\{1,\ldots,d\}$. All the $2^d$ terms in (\ref{eq:hoeffding}) have zero mean and are mutually uncorrelated with each other. This decomposition is unique and leads to the Sobol' indices. These are the elements of the $g(X)$ variance decomposition according to the different groups of input parameter interactions in (\ref{eq:hoeffding}). More precisely, for each $u\subset\{1,\ldots,d\}$, the first-order and total Sobol' indices of $X_u$ are defined by:
$$S_u=\frac{\mbox{Var}\left[g_u(X_u)\right]}{\mbox{Var}\left[g(X)\right]}\text{ and }S_u^T=\sum_{v\supset u}S_v.$$
$S_u$ represents the part of the output variance explained by $X_u$, independently from the other inputs, and $S_u^T$ is the part of the output variance explained by $X_u$ considered separately and in interaction with the other input parameters.

In practice, we are usually interested in the first-order sensitivity indices $S_1,\ldots,S_d$, the total ones $S_1^T,\ldots,S_d^T$ and sometimes in the second-order ones $S_{ij}$, $ 1\leq i<j \leq d$. The model $g$ is devoid of interactions if $\sum_{i=1}^dS_i\approx 1$.

\section{Appendix: Sobol' indices from a joint metamodel}\label{sec:soboljoint}

By adopting the same notations as in Section \ref{jointGp} where $\mathbf{X_{exp}}$ is the vector of all the controllable (or explanatory) inputs and $X_\varepsilon$ denotes the uncontrollable input (group of non explanatory inputs), we can show that the variance of the output variable $Y(\mathbf{X_{exp}},X_\varepsilon)$ can be rewritten and deduced from the two components $Y_m$ and $Y_d$ (equations (\ref{eqYm}) and (\ref{eqYd})):
\begin{equation}  \label{decompvarcond}
  \mbox{Var} [Y(\mathbf{X_{exp}},X_\varepsilon)] = \displaystyle \mbox{Var}_{\mathbf{X_{exp}}} \left[Y_m(\mathbf{X_{exp}}) \right] + \mathbb{E}_{\mathbf{X_{exp}}} \left[Y_d(\mathbf{X_{exp}}) \right] 
\end{equation}
where $\E_{X}$ (resp. $\mbox{Var}_{X}$) denotes the mean (resp. variance) operator with respect to the probability density function of $X$.
Furthermore, the variance of $Y$ is the sum of the contributions of both all the explanatory inputs in $\mathbf{X_{exp}}$ and $X_\varepsilon$:
\begin{equation}
\displaystyle  \mbox{Var}(Y) = V_{\varepsilon}(Y) + \sum_{u\subset\mathbf{X_{exp}}} [V_u(Y) + V_{u\varepsilon}(Y)]
\end{equation}
where $V_\varepsilon(Y)=\mbox{Var}_{X_\varepsilon}[\mathbb{E}_{\mathbf{X_{exp}}}(Y|X_\varepsilon)]$, 
$V_u(Y)=\mbox{Var}_{\mathbf{X_u}}[\mathbb{E}_{\mathbf{X_{-u}}}(Y|X_u)] -\sum_{v \subsetneq u}V_v(Y)$, $V_{u\varepsilon}(Y)=\mbox{Var}_{X_u X_\varepsilon}[\mathbb{E}_{\mathbf{X_{exp\; -u}}}(Y|X_u X_\varepsilon)]-V_u(Y)-V_\varepsilon(Y)$.

Variance of the mean component $Y_m(\mathbf{X_{exp}})$ denoted hereafter $Y_m$ can be also decomposed:
\begin{equation}
\displaystyle  \mbox{Var}(Y_m) = \sum_{u\subset\mathbf{X_{exp}}}  V_u(Y_m)\;.
\end{equation}

As $V_u(Y_m) = \mbox{Var}_{\mathbf{X_u}} \mathbb{E}_\mathbf{X_{exp \, -u}}  [\mathbb{E}_{X_\varepsilon} (Y|\mathbf{X_{exp}})|X_u] = V_u(Y)$, Sobol' indices according to any subset of input variables of $\mathbf{X_{exp}}$ can be derived and estimated from $Y_m$:
\begin{equation}\label{eqSiY} 
S_u =  \frac{V_u(Y_m)}{\mbox{Var}(Y)} \, \text{for any} \; u \subset \mathbf{X_{exp}}.
\end{equation}
Similarly, the total sensitivity index of $X_\varepsilon$ is given by:
\begin{equation}\label{eqSTeps} 
S_{\varepsilon}^{T} = \displaystyle \frac{V_\varepsilon(Y) + \sum_{u \subset \mathbf{X_{exp}}} V_{u\varepsilon}(Y)}{\mbox{Var}(Y)} = \frac{\mathbb{E}_{\mathbf{X_{exp}}}[Y_d(\mathbf{X_{exp}})]}{\mbox{Var}(Y)} \;. 
\end{equation} 
Note that, as $Y_d(\mathbf{X_{exp}})$ is a positive random variable, the positivity of $S_{\varepsilon}^{T}$ is guaranteed. 

\end{document}